\documentclass{article}
\pdfoutput=1

\usepackage{arxiv}
\usepackage[utf8]{inputenc}
\usepackage[T1]{fontenc}    
\usepackage[sort&compress,numbers]{natbib}

\usepackage{subfigure}
\usepackage{graphicx}

\usepackage{pstool}
\usepackage{psfrag}       

\usepackage{amsmath,bm,amsfonts,mathrsfs}   
\usepackage{color}

\usepackage{hyperref}       
\usepackage{url}            
\usepackage{enumitem}

\usepackage{shortcuts}

\newcommand{\bsym}[1]{\boldsymbol{#1}}

\usepackage{xcolor}
\definecolor{mintedbg}{gray}{0.95}
\usepackage{mdframed}

\usepackage{booktabs}
\usepackage{etoolbox}
\usepackage{siunitx}
\usepackage{multirow}

\DeclareFixedFont{\ttb}{T1}{txtt}{bx}{n}{12} 
\DeclareFixedFont{\ttm}{T1}{txtt}{m}{n}{12}  

\usepackage{color}
\definecolor{deepblue}{rgb}{0,0,0.5}
\definecolor{deepred}{rgb}{0.6,0,0}
\definecolor{deepgreen}{rgb}{0,0.5,0}

\usepackage{listings}

\newcommand\pythonstyle{
\lstset{
language=Python,
basicstyle=\small\ttfamily,
numberstyle=\color{gray},
stringstyle=\color[HTML]{933797},
commentstyle=\color[HTML]{228B22}\small\sffamily,
emph={[2]from,import,pass,return}, emphstyle={[2]\color[HTML]{DD52F0}},
emph={[3]range}, emphstyle={[3]\color[HTML]{D17032}},
emph={[4]for,in,def}, emphstyle={[4]\color{blue}},
showstringspaces=false,
breaklines=true,
prebreak=\mbox{{\color{gray}\tiny$\searrow$}},
xleftmargin=0pt
}
}

\lstnewenvironment{python}[1][]
{
\pythonstyle
\lstset{#1}
}
{}

\newcommand\pythoninline[1]{{\pythonstyle\lstinline!#1!}}

\newcommand\bashstyle{\lstset{
language=Bash,
basicstyle=\ttm\small,
otherkeywords={self},             
keywordstyle=\ttm,
showstringspaces=false            %
}}

\lstnewenvironment{bash}[1][]
{
\bashstyle
\lstset{#1}
}
{}

\newcommand\parameterstyle{\lstset{
language=XML,
basicstyle=\ttm\small,
otherkeywords={self},             
keywordstyle=\ttm,
showstringspaces=false            %
}}

\lstnewenvironment{parameter}[1][]
{
\parameterstyle
\lstset{#1}
}
{}

\def\BTheta{\mbox{\boldmath$\Theta$}}

\def\Btheta{\mbox{\boldmath$\theta$}}

\def\Bchi{\mbox{\boldmath$\chi$}}

\def\Bomega{\mbox{\boldmath$\omega$}}

\def\bR{\mbox{\boldmath$ R$}}

\def\bx{\mbox{\boldmath$ x$}}
\def\by{\mbox{\boldmath$ y$}}

\def\GTfield{\phi}
\def\GTChemPotential{M}
\def\GTFreeEnergy{\psi}
\def\Obsfield{\varphi}
\def\ObsFreeEnergy{\Psi}
\begin{document}

\title{\texttt{mechanoChemML}: A software library for machine learning in computational materials physics}
\author{X.~Zhang$^1$, G.H.~Teichert$^1$, Z.~Wang$^1$, M. Duschenes$^{2}$, S.~Srivastava$^1$, \\ E.~Livingston$^1$, J.~Holber$^2$, M.~Faghih Shojaei$^1$, A.~Sundararajan$^1$ and K.~Garikipati$^{1,2,3,4}$\thanks{Corresponding author. E-mail address: krishna@umich.edu}\\
$^1$Department of Mechanical Engineering, University of Michigan, United States \\
$^2$Applied Physics Program, University of Michigan, United States \\
$^3$Department of Mathematics, University of Michigan, United States \\
$^4$Michigan Institute for Computational Discovery \& Engineering, University of Michigan, United States}

\maketitle

\begin{abstract}
We present \texttt{mechanoChemML}, a machine learning software library for computational materials physics. \texttt{mechanoChemML} is designed to function as an interface between platforms that are widely used for  machine learning on one hand, and  others for solution of partial differential equations-based models of physics. Of special interest here, and the focus of \texttt{mechanoChemML}, are applications to computational materials physics. These typically feature the coupled solution of material transport, reaction, phase transformation, mechanics, heat transport and electrochemistry. Central to the organization of \texttt{mechanoChemML} are machine learning workflows that arise in the context of data-driven computational materials physics. The \texttt{mechanoChemML} code structure is described, the machine learning workflows are laid out and their application to the solution of several  problems in materials physics is outlined.
\end{abstract}

\keywords{machine learning software library \and machine learning workflows \and computational materials physics \and partial differential equation solvers \and scientific software}

\section{Introduction}
\label{sec:intro}
Until roughly a decade ago, the dominant theme in computational materials physics was the forward solution of a very wide array of problems by using methods that ranged from electronic structure calculations, through molecular and statistical mechanics computations, to partial differential equations (PDEs) describing continuum phenomena. Data, while used, was rarely a part of the computational workflow. Much has changed since then with the rise of of data-driven modeling and machine learning (ML) in particular. It is now routine for computations with each of the above classes of methods to ingest data, produce and employ them as the packets of communication with other simulations, or experimental platforms. Thus, Density Functional Theory (DFT), long the workhorse of electronic structure calculations for materials applications, may now be based on full field \emph{ab initio} solutions from Configuration Interaction methods to inform or learn the exchange correlation functional \cite{kanungo2019exact}. Formation energy data computed by DFT parameterize interatomic potentials for molecular dynamics \cite{wang2019coarse} or, through cluster Hamiltonians, drive Monte Carlo (MC)-based statistical mechanics simulations \cite{Natarajan2017}. And, all of these methods generate data for continuum scale PDEs \cite{Teichert2018,Teichert2019,Teichert2020}. The data and information flow also travels down the scales as, for example, strain fields drive deformation potentials in DFT.

Notably, the interaction between the model formalisms at different levels in this outline of scale bridging in materials physics is not a trivial matter. Efficient representations are needed, often in the form of high-dimensional, or conversely, reduced dimensionality functions or functionals. For instance, MC simulations of phase or order-disorder transitions yield millions of configurations from which an $\mathcal{O}(10)$-dimensional free energy surface needs to be parameterized for continuum phase field or elasticity computations \cite{Teichert2020,teichert2021li}. This is an obvious entry  point for neural network-based ML methods. The continuum simulations now commonly entail billions of degrees of freedom evolving over time scales ranging from microseconds to hours. The imperative of high-throughput design and material optimization problems needs reduced-order models ranging from homogenization methods through neural networks of various flavors \cite{frankel2019predicting,wang2018multiscale,le2015computational,Zhang2020Garikipati-CMAME-ML-RVE} and modern optimization methods to graph theoretic representations \cite{duschenes2021reduced}. Also of very recent interest are physics-constrained machine learning techniques that ensure fidelity of fast solutions of continuum PDEs \cite{Zhang2021Garikipati-BNN-weak-solution-PDE-SS,Zhu2018Zabaras-UQ-Bayesian-ED-CNN,Zhu2019Perdikaris-Physics-PDE-CNN,Winovich2019Lin-ConvPDE-UQ,Bhatnagar2019CNN-encoder-decoder,Li2020Reaction-diffusion-prediction-CNN,Lagaris1998NN-PDEs,Han2018Solve-high-dimension-PDE-deep-learning,Sirignano2018DGM-solve-PDEs,Raissi2019physics-informed-forward-inverse-jcp,Geneva2020Zabaras-JCP-auto-regressive-NN-PDE,Yang2021BPINNs-PDE,Berg2018unified-deep-ann-PDE-complex-geometries}.

The reliance on data ingestion, transfer, high-dimensional as well as reduced-order function representations naturally rests upon the impressive, robust and rapidly growing ecosystem for machine learning, and more broadly artificial intelligence and even all of data science. However, there are aspects specific to data-driven computational materials physics that have led to a sub-ecosystem of mathematical techniques, algorithms and frameworks. This sub-ecosystem is now spawning software that interfaces platforms such as \texttt{TensorFlow} \cite{abadi2016tensorflow}, \texttt{Keras} \cite{chollet2015keras}, \texttt{PyTorch} \cite{paszke2019pytorch} and their ilk with the foundational software for scientific computing such as \texttt{PETSc} \cite{balay2019petsc}, \texttt{Trilinos} \cite{heroux2005overview} and yet others built on top of them such as \texttt{deal.ii} \cite{alzetta2018deal}, \texttt{FEniCS} \cite{alnaes2015fenics} and even their derivatives. The \texttt{mechanoChemML} library belongs in this sub-ecosystem.

\section{Background to the methods}
\label{sec:backmethods}

\begin{figure}
\centering
\includegraphics[width=1.0\linewidth]{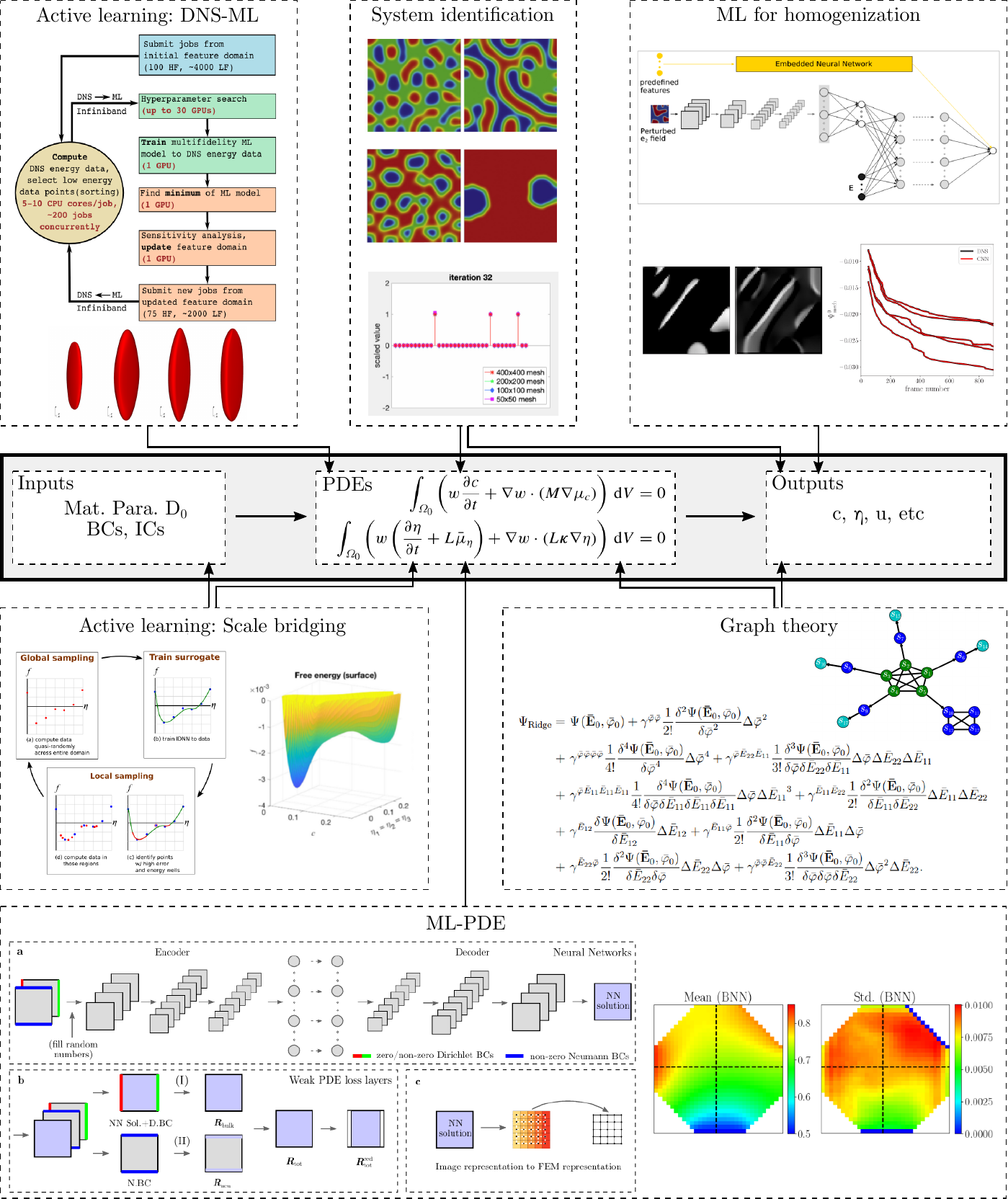}
\caption{A schematic illustrating the range of ML methods comprising \texttt{mechanoChemML} for data-driven computational material physics.}
\label{fig:summary-figure}
\end{figure}

\texttt{mechanoChemML} leverages PDE solver libraries such as \texttt{deal.ii, FEniCS} \cite{alzetta2018deal,alnaes2015fenics} \cite{mechanoChem_github} and others built on them, such as \texttt{mechanoChem}, also developed by the authors. These libraries present frameworks with varying degrees of abstraction for the solution of PDEs. They exploit the range of high-performance computing methods including geometry and mesh generation, vectorization, automatic differentiation, linear and nonlinear solvers, threading and parallelism, among others. With the availability of cluster computing resources, both fixed and in the cloud, this has made the large scale solution of multiphysics problems commonplace and widely accessible. As one natural next step, this ability to compute PDE solutions, almost at will, has led to thinking about enhancing the fidelity and speed of modeling as well as further exploiting the solutions. We introduce examples of such thinking that has led to \texttt{mechanoChemML}.

Phase field modeling is centered on a parabolic PDE for the evolution of a scalar or vector variable that parameterizes a free energy density function. The dissipative dynamics is governed by minimization of the free energy. In many situations, the final (near-)equilibrium state is of greater interest than the dynamics itself. This leads to alternate views of the problem as one of non-convex optimization over the free energy density landscape. Since the optimization entails computation of feasible solutions in large numbers, it suggests neural network surrogates for the free energy density in terms of reduced-order representations. Multi-fidelity learning, sensitivity analysis and a wrapper of active learning help improve the efficiency of the surrogate optimization \cite{Teichert2018}.

Scale bridging computational frameworks have long been of interest in materials physics. A peculiarity of this problem is that the model description changes in a fundamental manner across scales: from electronic structure computations, typically DFT, to a variety of Monte Carlo simulations of molecular configurations to recreate the statistical mechanics, through continuum PDEs of elastic or inelastic behavior. Central quantities across these scales are the free energy, or its density and derivatives (e.g., chemical potentials and stresses). Their fundamental parameterization comes from sparse DFT computations. Their upscaling through Monte Carlo  is dependent on surrogate representations via machine learning methods. The next step in bridging scales--up to the continuum--brings special considerations of integrability as well as high-dimensional representation, both of which we have found convenient to satisfy with a breed of specially developed neural networks. Monte Carlo sampling in regions of interest also naturally has invoked active learning techniques \cite{Teichert2019, Teichert2020}.

At the continuum scale, a class of PDE models that combine phase field and non-convex elasticity underlies a wide range of problems that develop finely detailed microstructure. The effective mechanical response of these microstructures is obtained by homogenization methods, which are reduced-order models of a type. It has been natural to turn to deep and convolutional neural networks (DNNs and CNNs) for this problem. While the former require the insightful selection of features with the proper invariances, CNNs with an auto encoder-decoder structure afford the flexibility of learning from microstructures as images. This is the first example in which the imposition of physical constraints--of invariances and simpler symmetries--has arisen in our work \cite{Zhang2020Garikipati-CMAME-ML-RVE}.

The high-throughput solution of PDEs for inverse modeling, design and optimization leads to requirements of very fast solutions that are largely beyond the capability of traditional PDE solver libraries, such as those introduced at the beginning of this section. The development of PDE solvers is one of the most rapidly advancing fields in scientific machine learning. Neural networks have been a natural choice, with the added imperative of embedding the physics of the PDE as constraints. With field solutions regarded as images, CNNs in auto encoder-decoder structures are a natural choice, and the extension to uncertainty quantification made possible by Bayesian neural networks. This is the first instance in which learning from label-free data has arisen in our work \cite{Zhang2021Garikipati-BNN-weak-solution-PDE-SS}.

Inference of PDEs from data--inverse modeling of not just parameters but of the entire set of algebraic and differential operators is feasible with the availability of extensive data, regression, and more generally, nonlinear optimization methods. Along with the development of neural network PDE solvers, system inference enjoys advantages when the variational setting of the weak form, as well as discretization structures such as finite elements and finite differences are exploited. A number of supporting methods become of relevance for this problem, including those for sparse and noisy data, and gradient optimization \cite{WangCMAME2019, WangCMAME2021, Wangetal-COVID2020, Wangetal-COVID2021, Wang2020-MRu}.

Finally, reduced-order modeling on a physical systems scale is feasible with large datasets of computed solution states. We have exploited a  correspondence between the properties dictated by PDEs on solution states of physical systems and graph theory \cite{Banerjee2019}. In addition to representation and analysis on the systems scale by exploiting the machinery of graph theory, this opens the door to reduced-order modeling by exploiting a nonlocal calculus \cite{duschenes2021reduced}. It leverages methods developed for system inference, mainly stepwise regression to choose the reduced-order model.

\texttt{mechanoChemML} addresses the above ecosystem of machine learning techniques, as illustrated in Fig. \ref{fig:summary-figure}. Section \ref{sec:structure} discusses the structure of the library. Some of the above scientific applications are presented as examples using the \texttt{mechanoChemML} library in Section \ref{sec:sciexamples}. These two sections are the mainstay of the paper. {Installation, documentation, and contributing} is discussed in Section \ref{sec:install}, and closing remarks appear in Section \ref{sec:concl}.

\section{Structure of the \texttt{mechanoChemML} library}
\label{sec:structure}
\begin{figure}
\centering
\includegraphics[width=1.0\linewidth]{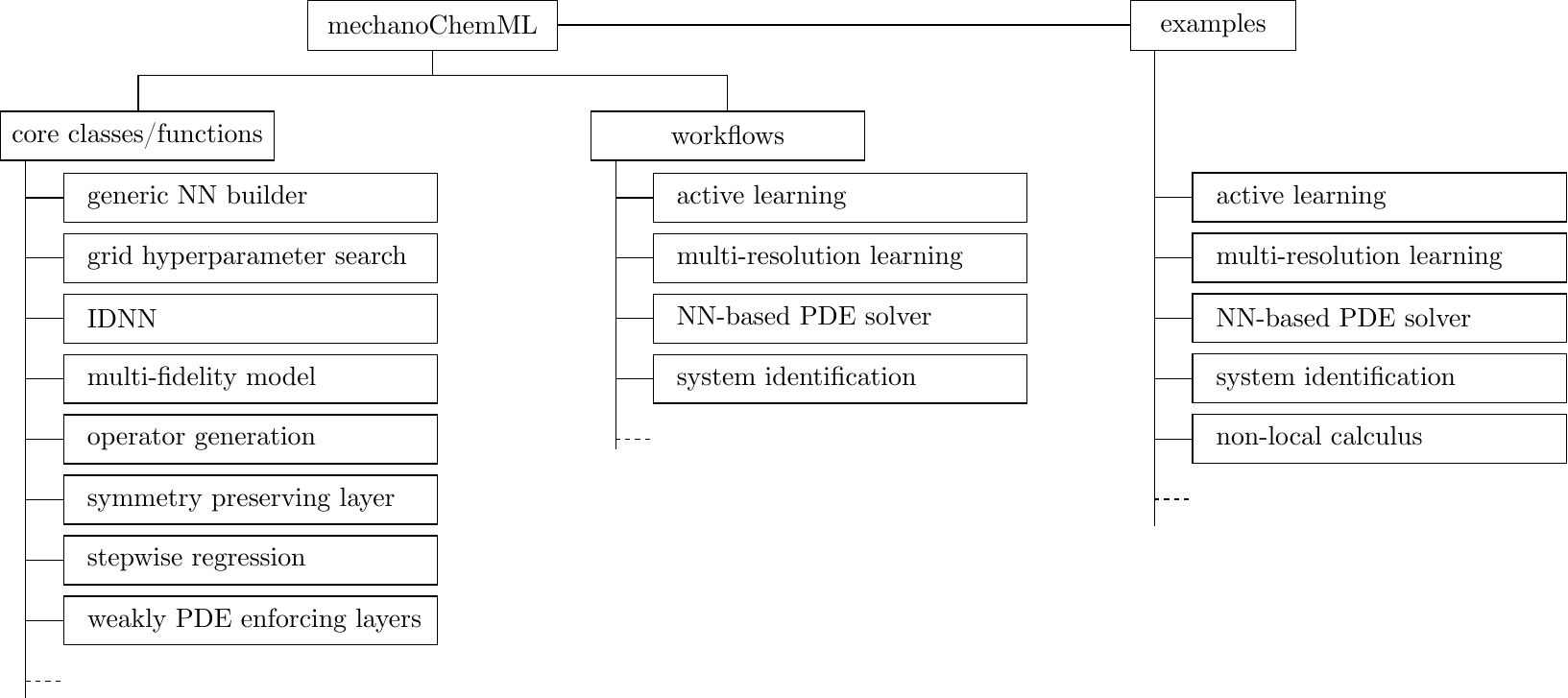}
\caption{Illustration of the structure and components of the \texttt{mechanoChemML} library {and the provided examples.}}
\label{fig:universe}
\end{figure}

The \texttt{mechanoChemML} library can be divided into two parts: the machine learning (ML) class library and example workflows (see Figure \ref{fig:universe}). {Scientific examples using the \texttt{mechanoChemML} library are discussed in Section \ref{sec:sciexamples}. }

\subsection{The machine learning (ML) class library}
The ML class library is a collection of classes and functions that are related to neural networks or other machine learning methods and are useful in data-driven modeling of mechanochemistry but do not already exist in well-used machine learning libraries. These classes and functions are in most cases, however, built on existing libraries. A list of the neural network architectures/layers and machine learning related algorithms appears in Figure \ref{fig:universe} and examples of their use in select applications are given in section \ref{sec:workflows}.

\subsubsection{Integrable deep neural network}
\label{sec:idnn}

\begin{figure}[tb]
\centering
\includegraphics[width=0.6\linewidth]{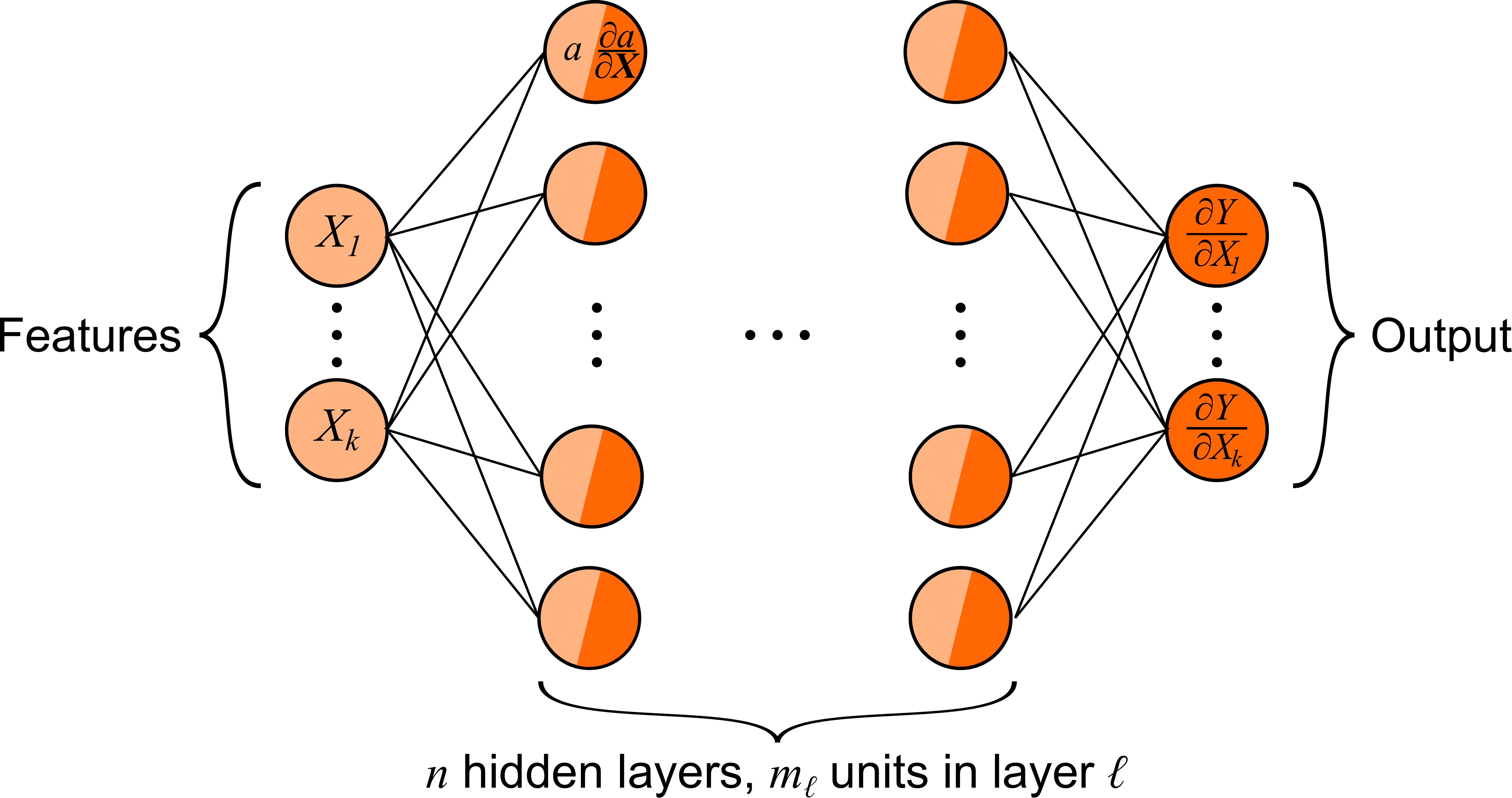}
\caption{A schematic of an IDNN, which is trained to derivative data and can be analytically integrated to recover the antiderivative function.}
\label{fig:IDNN}
\end{figure}

Integrable deep neural networks (IDNNs) \cite{Teichert2019,Teichert2020} are a type of neural network structure that allows training to derivative data and analytic integration to recover a neural network representation of the antiderivative. A schematic is shown in Figure \ref{fig:IDNN}. The IDNN is constructed as the derivative of a standard DNN so that after an IDNN is trained to the derivative data, its antiderivative is simply the standard DNN with the IDNN's trained weights and bias.

A standard DNN with $n$ hidden layers, where $\bsym{W}_\ell$, $\bsym{b}_\ell$ are the weight matrix and bias vector of hidden layer $\ell$, $g$ is the activation function, $a_\ell$ and $z_\ell$ are intermediate vector values at each layer, and $Y$ is the DNN output, can be represented with the following:
\begin{equation}
\begin{split}
\bsym{z}_\ell &= \bsym{b}_\ell + \bsym{W}_\ell\bsym{a}_{\ell-1}\\
\bsym{a}_\ell &= g(\bsym{z}_\ell)\\
Y &= \bsym{b}_{n+1} + \bsym{W}_{n+1}\bsym{a}_{n}
\end{split}
\end{equation}
Differentiation of $Y$ leads to the following additional equations that describe the IDNN, which is represented by $\partial Y/\partial X_k$:
\begin{equation}
\begin{split}
\frac{\partial \bsym{a}_\ell}{\partial X_k} &= g'(\bsym{z}_\ell)\odot\left(\bsym{W}_\ell\frac{\partial \bsym{a}_{\ell-1}}{\partial X_k}\right)\\
\frac{\partial Y}{\partial X_k} &= \bsym{W}_{n+1}\frac{\partial \bsym{a}_n}{\partial X_k}
\end{split} \label{eq:idnn}
\end{equation}
where the operator $\odot$ denotes element-wise multiplication. Note that both the activation function and its derivative are used in the IDNN.

Instead of directly implementing Eq. (\ref{eq:idnn}) to train an IDNN, however, we take advantage of the ability of deep learning libraries to apply a gradient operator to a standard neural network structure. To illustrate this approach, we consider a standard DNN to be represented as a function $\bsym{Y}(\bsym{X},\bsym{W},\bsym{b})$ of inputs $\bsym{X}$, weights $\bsym{W}$, and biases $\bsym{b}$. Then, the training of an IDNN to a set of first derivative data  $\{(\bsym{\hat{X}}_\theta,\bsym{\hat{y}}_{k_\theta})\}$ is the minimization of the mean square error of the gradient of a standard DNN (i.e. the IDNN) and the derivative data over the space of weights and biases:
\begin{align}
\bsym{\hat{W}},\bsym{\hat{b}} = \underset{\bsym{W},\bsym{b}}{\mathrm{arg\,min}}\,\sum_{k=1}^n\mathrm{MSE}\left(\frac{\partial\bsym{Y}(\bsym{X},\bsym{W},\bsym{b})}{\partial X_k}\Big |_{\bsym{\hat{X}}_\theta},\hat{y}_{k_\theta}\right)
\label{eq:IDNN-Wb}
\end{align}

The \texttt{mechanoChemML} library provides an \texttt{idnn} module that returns an IDNN implemented in \texttt{Keras} and capable of being trained on first derivative (gradient) data, second derivative (Hessian) data, and/or data from the function itself, provided at
\begin{mdframed}[backgroundcolor=mintedbg, linecolor=mintedbg, innerleftmargin=0, innertopmargin=0,innerbottommargin=0]
\begin{python}
mechanoChemML.src.idnn.IDNN
\end{python}
\end{mdframed}
The following example code shows how an IDNN with four inputs and two hidden layers of 20 neurons each would be initialized and trained to first derivative data:
\begin{mdframed}[backgroundcolor=mintedbg, linecolor=mintedbg, innerleftmargin=0, innertopmargin=0,innerbottommargin=0]
\begin{python}
idnn = IDNN(4,[20,20])

idnn.compile(loss=[None,'mse',None],
optimizer=keras.optimizers.RMSprop(lr=0.01))

idnn.fit(c_train,
mu_train,
epochs=1000,
batch_size=20)
\end{python}
\end{mdframed}

A test function for the IDNN model is included in the test for one of the utility functions and is briefly described in Section \ref{sec:convexity}. The IDNN is used in an example active learning workflow in Section \ref{sec:free_energy} to train a free energy density function where the training data are derivatives of the free energy.

\subsubsection{IDNN convexity}
\label{sec:convexity}

Since the IDNN was created to represent energy data, it is useful to determine if a point or set of points lies in a well (i.e., convex region) of the IDNN. The library includes functions that report the convexity of one or more points by checking if the Hessian, returned by the IDNN model, is positive definite. These functions are included at
\begin{mdframed}[backgroundcolor=mintedbg, linecolor=mintedbg, innerleftmargin=0, innertopmargin=0,innerbottommargin=0]
\begin{python}
mechanoChemML.src.idnn
\end{python}
\end{mdframed}
An example using the IDNN convexity functions is shown in Figure \ref{fig:active_learning} of Section \ref{sec:active_learning}, which describes active learning workflows. If desired, additional characteristics of the IDNN could also be evaluated for use in an active learning or other workflow.

A test function was written to confirm that the convexity functions and the IDNN are performing the differentiation and positive definiteness check correctly. This is done by comparing the code's result on an IDNN with known weights to the correct results. The test is located at
\begin{mdframed}[backgroundcolor=mintedbg, linecolor=mintedbg, innerleftmargin=0, innertopmargin=0,innerbottommargin=0]
\begin{python}
mechanoChemML.testing.idnn_convexity_test
\end{python}
\end{mdframed}

\subsubsection{Transform layer}
Sometimes it is desirable to apply a transform to the inputs of a neural network. In addition to the relatively simple tasks of shifting and scaling data, this is one way to enforce some types of symmetry. For example, if a neural network should be symmetric about an input $x$, the network can use the transform $x^2$ as the input instead of $x$ itself. Higher dimensional symmetries, such as rotational symmetries, would include symmetry preserving functions of multiple input variables. We provide a custom transform layer using \texttt{Keras} that allows the user to specify a function, \texttt{transforms}, that takes an input array \texttt{x} and returns a list of the transformed outputs, as in the following example snippet:
\begin{mdframed}[backgroundcolor=mintedbg, linecolor=mintedbg, innerleftmargin=0, innertopmargin=0,innerbottommargin=0]
\begin{python}
def transforms(x):
return [x[0],x[1]**2]

y = Transform(transforms)(x)
\end{python}
\end{mdframed}
In this example, the first variable is taken as is, but the second variable is squared to enforce a symmetry about $x_1 = 0$. The transform layer is used in the construction of the \texttt{IDNN} model (see Section \ref{sec:idnn}) to allow symmetry enforcement and is provided at the following location:
\begin{mdframed}[backgroundcolor=mintedbg, linecolor=mintedbg, innerleftmargin=0, innertopmargin=0,innerbottommargin=0]
\begin{python}
mechanoChemML.src.transform_layer.Transform
\end{python}
\end{mdframed}

\subsubsection{General neural network creation}
Data-driven modeling research often requires a hyperparameter study, which involves the exploration of different NN architectures.
To facilitate it, a collection of general NN classes, such as \verb NN_user_general , \verb BNN_user_general , and \verb BNN_user_weak_pde_general , are provided in \verb mechanoChemML.src.nn_models ~to construct NNs based on inputs from a configuration file.
Such classes offer the flexibility to explore different NN architectures, particularly complex ones.
They also allow non-expert users to quickly setup different NNs.
For example, the following input will construct a deterministic NN with an encoder-decoder structure.
\begin{mdframed}[backgroundcolor=mintedbg, linecolor=mintedbg, innerleftmargin=0, innertopmargin=0,innerbottommargin=0]
\begin{parameter}
NNArchitecture =
type=PDERandom;
type=Conv2D | filters=8 | kernel_size=5 | activation=relu | padding=same;
type=MaxPooling2D | pool_size=(2,2) | padding=same;
type=Conv2D | filters=8 | kernel_size=5 | activation=relu | padding=same;
type=MaxPooling2D | pool_size=(2,2) | padding=same;
type=Flatten;
type=Dense | units=64 | activation=relu;
type=Reshape | target_shape=[4,4,4];
type=Conv2D | filters=8 | kernel_size=5 | activation=relu | padding=same;
type=UpSampling2D | size=(2,2);
type=Conv2D | filters=8 | kernel_size=5 | activation=relu | padding=same;
type=Conv2D | filters=1 | kernel_size=5 | activation=linear | padding=same;
\end{parameter}
\end{mdframed}
One can also easily change the layer types in the configuration file to names of their corresponding probabilistic implementation in the \texttt{TensorFlow Probability} library to construct a probabilistic NN with identical architecture.

\subsubsection{Weak PDE enforcing layers}
\begin{figure}[h!]
\centering
\includegraphics[width=1.0\linewidth]{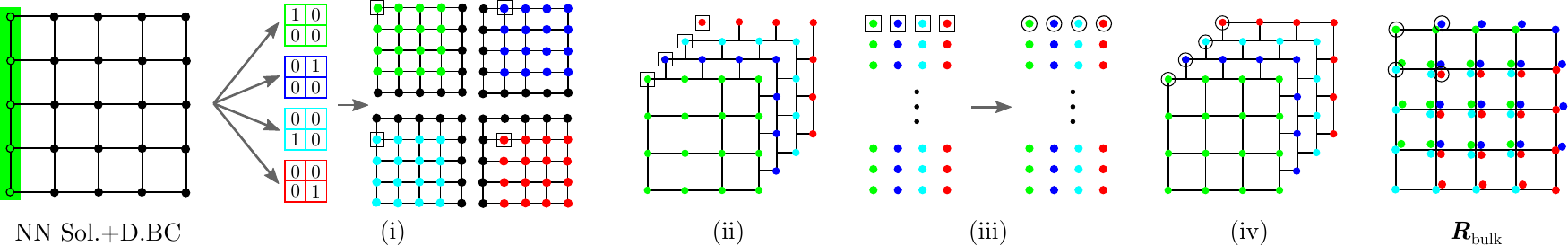}
\caption{Illustration of steps to compute the bulk residual $\bR_\text{bulk}$ based on the NN inputs (BC info) and NN outputs through different convolutional operator-based, and vectorized calculations.}
\label{fig:pde-layer-ill}
\end{figure}

A major thrust in using machine learning techniques of interest in the computational mechanics, materials, and physics communities is to solve PDEs with little or no pre-labeled data.
In the \texttt{mechanoChemML} library, a collection of classes, layers, and functions, such as,
\begin{mdframed}[backgroundcolor=mintedbg, linecolor=mintedbg, innerleftmargin=0, innertopmargin=0,innerbottommargin=0]
\begin{python}
mechanoChemML.src.pde_layers.GetElementResidualMask
mechanoChemML.src.pde_layers.ComputeBoundaryMaskNodalData
mechanoChemML.src.pde_layers.ComputeNeumannBoundaryResidualNodalData
mechanoChemML.src.pde_layers.Get1DGaussPointInfo
mechanoChemML.src.pde_layers.Get2DGaussPointInfo
mechanoChemML.src.pde_layers.GetNodalInfoFromElementInfo
mechanoChemML.src.pde_layers.LayerBulkResidual
\end{python}
\end{mdframed}
are provided in \verb mechanoChemML.src.pde_layers ~to compute the discretized residual, as illustrated in Fig. \ref{fig:pde-layer-ill}, through an efficient, convolutional operator-based, and vectorized residual calculation implementation based on the framework proposed in \cite{Zhang2021Garikipati-BNN-weak-solution-PDE-SS}. The listed classes, layers, and functions will be used in the workflow related to the NN-based PDE solver that is discussed in Section \ref{sec:workflow-pde}.

\subsubsection{Non-local calculus on graphs}
Graph-based representation methods have shown great success in reduced-order modeling for a range of physical phenomena \cite{banerjee2019graph}, that often lend themselves to a PDE form. Non-local calculus on graphs (see \cite{Gilboa2008}) provide a mathematical framework for defining and evaluating these PDE models. A graph intrinsically allows representation of unstructured data, making it suitable for data-driven system identification when the data is sparse and non-uniformly distributed. For this reason, it has been applied to a variety of problems   \cite{Gilboa2008,Desquesnes2013,Hein2007}.

A graph, denoted by $G = (V,E)$, consists of a set of vertices, $i \in V$, with $\vert V \vert = n$. The vertices are connected by a set of edges, $e \in E$, where each edge $e = (i,j)$ is a pair of vertices. A data set provided in the form, $\lbrace{ (\boldsymbol{x}_1, u_1), \cdots, (\boldsymbol{x}_n, u_n) \rbrace}$, with independent variables, $\boldsymbol{x} \equiv [x^1, \cdots x^p] \in \Omega \subset \mathbb{R}^p$,
and dependent variables, $u:\Omega \rightarrow \mathbb{R}$, can be represented as a graph, $(V,E)$ by placing a vertex of the graph at $\boldsymbol{x}_i$ for each data point. Furthermore, the function(al) values, $u_i \equiv u(\boldsymbol{x}_i)$ are treated as the states of the of the vertex. The edges on the graph can then be formed between points that lie in each other's local neighborhoods, $\mathcal{N}(\boldsymbol{x})$.

The graph can also have global and local attributes on the vertices and edges, as we describe below. Of importance are edge weights $w$, that are generally functions of the local vertex attributes, and admit a notion of distance between states on the graph. Scalars, $u(\boldsymbol{x}_{i})$ correspond to function(al)s at the $i^{th}$ vertex. Vectors $v(\boldsymbol{x}_{i},\boldsymbol{x}_{j})$ are functions on vertex pairs $i,j$ or the edge $e = (i,j)$. Gilboa \emph{et al.}~ define a discrete calculus, consisting of non-local operators, based on differences between states of vertices $i,j \in V$, and an edge weight $w(\boldsymbol{x}_{i},\boldsymbol{x}_{j})$ \cite{Gilboa2008};

\begin{align}
\frac{\delta u }{\delta x^\mu } ( \widetilde{\boldsymbol{x}})= \sum_{\boldsymbol{x}\in \mathcal{N}(\widetilde{\boldsymbol{x}})} \frac{u(\boldsymbol{x}) - u(\widetilde{\boldsymbol{x}})}{z^\mu} w(\boldsymbol{x}, \widetilde{\boldsymbol{x}}), \qquad z= \boldsymbol{x}-  \widetilde{\boldsymbol{x}}
\end{align}

We recently developed a methodology to choose an optimal set of edges, $E$ and corresponding weights, $w$, to achieve arbitrary accuracy in the computation of these derivatives. The theoretical grounds for understanding the convergence of derivatives in non-local calculus to the standard definitions of local derivatives in terms of $h^p$-convergence in grid-based graphs are presented in Ref. \cite{duschenes2021reduced}. The non-local derivatives are computed in the \texttt{mechanoChemML} library's \texttt{graph} module. The user can access the main functionality of the class by calling the following function, that executes the various functions involved in the library.

\begin{mdframed}[backgroundcolor=mintedbg, linecolor=mintedbg, innerleftmargin=0, innertopmargin=0,innerbottommargin=0]
\begin{python}
mechanoChemML.src.graph_main.main
\end{python}
\end{mdframed}

The library handles Input/Output using \textit{comma-separated values (CSV)} files. The main function takes only one argument: the \emph{settings} dictionary, where the user defines the paths to input/output files, the model settings (e.g. dimension, $p$, of $\Omega$) and the set of operations on data (e.g. calculation of partial derivatives). The user can manipulate the accuracy and the neighborhood selection strategy for derivatives using nested dictionaries in settings.

The following example code shows how to read data, $(x^1, x^2, x^3, u^1, u^2)$, formatted with column names \verb ['x_1','x_2','x_3','u_1','u_2']  located in the file, \verb './data/func_val.csv' . First, the code demonstrates computation of an algebraic term, $u^3 = x^1 + x^2 + x^3$. Then estimation of a partial derivative $\delta u^3/\delta x^1$ is demonstrated. After the main function is executed, the new data is saved in \verb './result/data.csv' .

\begin{mdframed}[backgroundcolor=mintedbg, linecolor=mintedbg, innerleftmargin=0, innertopmargin=0,innerbottommargin=0]
\begin{python}
settings = {
#Path settings
'cwd': '.',
'directories_load':'data',
'directories_dump':'result',
'data_filename':'func_val.csv',

#Model settings
'model_order':2,
'model_p':3,

#Operations settings
'algebraic_operations': [[
{'func':lambda df: df['x_1'] + df['x_2'] + df['x_3'], 'labels':'u_3'} ]],

'differential_operations':[[
{'function':'u_3',
'variable': ['x_1'],
'weight':['stencil'],
'adjacency':['nearest'],
'manifold':[['x_1', 'x_2', 'x_3']],
'accuracy': [2],
'dimension':[0],
'order':1,
'operation':['partial']}]]}
mechanoChemML.src.graph_main.main(settings = settings)
\end{python}
\end{mdframed}

\subsection{ML workflows}
\label{sec:workflows}
The bulk of the library consists of several example workflows for specific applications. Each of these workflows combines various elements of the ML class library  with additional algorithms and methods to solve a problem of interest in computational materials physics. Introductory descriptions of the physics and the library components used in the workflows are presented in the following section.

\subsubsection{Active learning}
\label{sec:active_learning}
Deep learning can be used to create surrogate models for accurate-yet-expensive simulations. In these cases, the data used for training comes from computations. This provides a challenge and an opportunity: the challenge is that we must first compute the data we use to train, and the opportunity is that we can choose what data to compute. Since it would be infeasible, particularly in the case of a high-dimensional input space, to densely sample data uniformly across the entire feature space, we turn to active learning. Active learning methods interweave machine learning training with data sampling in a way that lets the machine learning model choose the data that will be most informative \cite{Settles2012}. This allows us to reduce the amount of data needed to train a sufficiently representative surrogate model.

\begin{figure}
\centering
\includegraphics[width=0.8\textwidth]{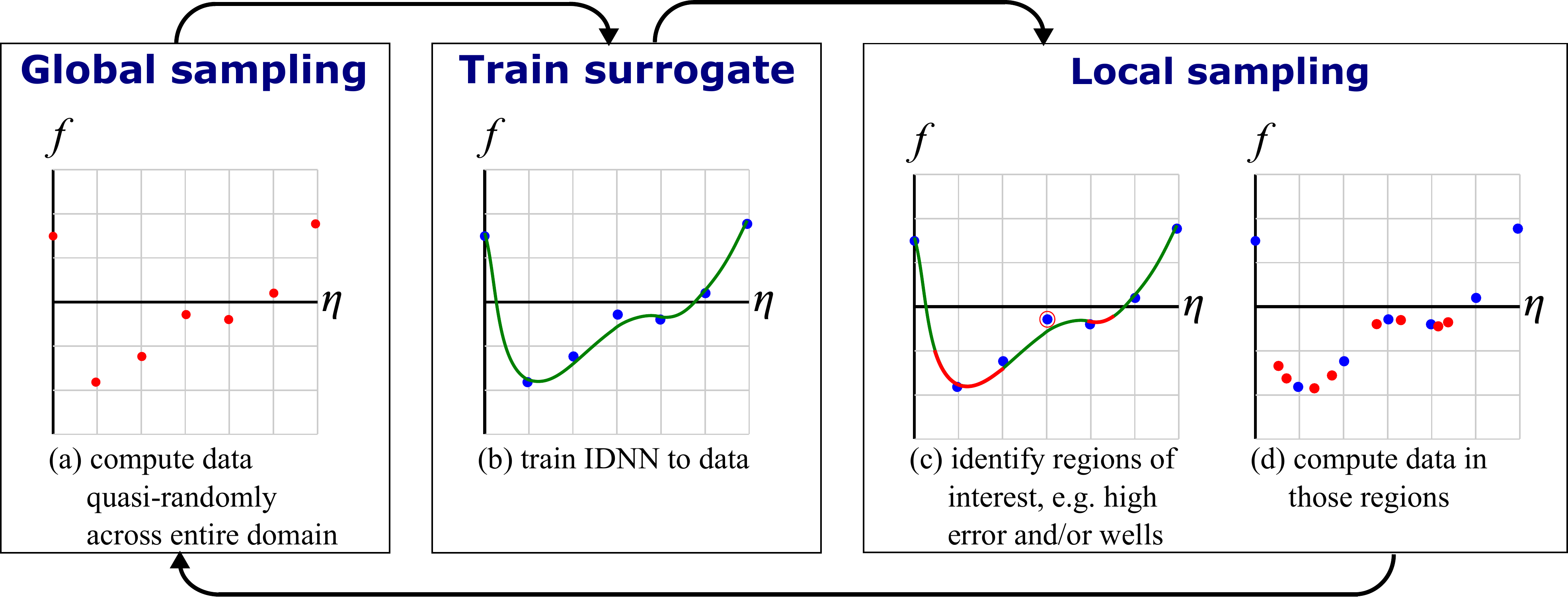}
\caption{Flowchart describing the active learning process with hypothetical 1D data.}
\label{fig:active_learning}
\end{figure}

The active learning workflows in the \texttt{mechanoChemML} library follow a general cycle of global sampling or exploration, training, and local sampling or exploitation. These are defined in the following functions
\begin{itemize}
\item \verb global_sampling ~
\item \verb surrogate_training ~
\item \verb local_sampling ~
\end{itemize}
from the active learning examples provided at
\begin{mdframed}[backgroundcolor=mintedbg, linecolor=mintedbg, innerleftmargin=0, innertopmargin=0,innerbottommargin=0]
\begin{python}
mechanoChemML.workflows.active_learning
\end{python}
\end{mdframed}
Global sampling or exploration allows data to be found in potentially interesting yet unexplored regions of the feature space, and it is also necessary to generate an initial training set. Training is performed using data sampled up to that point in the workflow, potentially down-selected based on a desired criterion. Active learning takes place during local sampling or exploitation, in which the currently trained network is queried to determine regions of interest where more data could be useful. This might simply determine areas where there is still a high pointwise error between the data and the neural network prediction. Local sampling might also be guided by the landscape of the DNN, such as convexity or high gradients. Other criteria can be used for local sampling, based on the important physics in the application. A flowchart describing the active learning process is shown in Figure \ref{fig:active_learning}.

The active learning workflow is demonstrated through an example scientific application in Section \ref{sec:free_energy}, which describes learning a high-dimensional free energy density function from first-principles data.

\paragraph{Testing} Since an active learning workflow is largely application specific, there is no generic test function for the workflow as a whole. However, it is recommended that the user confirm that a modified workflow is running correctly by using a set of smaller workflow parameters (e.g., fewer training iterations, reduced data generation, etc.) before running the full workflow.

\subsubsection{Stepwise regression for system identification}
\label{sec:stepwise}
We have developed a class of inverse modeling techniques, referred as as Variational System Identification (VSI) to allow the identification of physics from data \cite{WangCMAME2019,WangCMAME2021,WangWu2020}. It aims to infer the physics governing observed phenomena by identifying the minimal set of operators whose combination into PDEs completely describes the data. VSI has been applied to various real-world problems such as discovering physics from dynamical systems\cite{WangCMAME2019,WangCMAME2021}, identifying constitutive models of biological tissues\cite{Wang2020-MRu}, and inferring governing mechanisms in developmental biology and disease propagation\cite{Wang2020-brain,Wangetal-COVID2020,Wangetal-COVID2021}. VSI leverages the weak form unlike most other methods that use the strong form for identification of governing equations \cite{KutzPNAS2015,KutzIEEE2016,KutzSCIADV2017,KutzHybrid2018,KutzChaos2018,KutzCDC2018,KutzSIAM2019,kanungo2019exact}. It inherits many advantages from the weak form, including (a) the natural identification of boundary conditions, (b) introduction of basis functions to interpolate the data, which can be chosen to have high regularity, (c) transfer of derivatives to variations, thus lowering the regularity required in the data, and (d) isolating operators on the data by judicious choice of variations.

As a first step of data collection, users can import from their own data set. However, the \verb mechanoChemML ~library also provides an example DNS interface \verb mechanoChemML.third_party.dns_wrapper ~which can be extended to link with PDE solver packages such as \texttt{deal.ii} and \texttt{FEniCS}. With this interface, users can build a full pipeline of data generation to system identification. An example of the forward model to generate the patterns in Figure \ref{fig:patterns} can be found at
\begin{mdframed}[backgroundcolor=mintedbg, linecolor=mintedbg, innerleftmargin=0, innertopmargin=0,innerbottommargin=0]
\begin{python}
mechanoChemML.workflows.systemID.forward_model
\end{python}
\end{mdframed}

The \verb mechanoChemML ~library provides an operator construction module for constructing operators in weak form with the basis function from two families: polynomial basis functions traditional to finite element analysis (FEA), and the Non-Uniform Rational B-Splines (NURBS) used widely in Isogeomeric Analysis (IGA). A example of constructing operators in weak form as shown in Equations (\ref{eq:basis_weak}) and (\ref{eq:basis_Ct}) can be found at
\begin{mdframed}[backgroundcolor=mintedbg, linecolor=mintedbg, innerleftmargin=0, innertopmargin=0,innerbottommargin=0]
\begin{python}
mechanoChemML.workflows.systemID.generate_basis
\end{python}
\end{mdframed}

The approach to inference combines system identification by stepwise regression \cite{WangCMAME2019,WangWu2020} with a statistical criterion called the $F$-test for eliminating basis terms. This forms the main component in the stepwise regression module:
\begin{mdframed}[backgroundcolor=mintedbg, linecolor=mintedbg, innerleftmargin=0, innertopmargin=0,innerbottommargin=0]
\begin{python}
mechanoChemML.src.stepwiseRegression.stepwiseR
\end{python}
\end{mdframed}
The stepwise regression module provides two basis elimination strategies and a variety of regularization schemes for regression. Users may adopt the stepwise regression module and build their own system identification framework. However we have found that in most cases, users can set up their inference problems using a higher level module for system identification:
\begin{mdframed}[backgroundcolor=mintedbg, linecolor=mintedbg, innerleftmargin=0, innertopmargin=0,innerbottommargin=0]
\begin{python}
mechanoChemML.workflows.systemID.systemID
\end{python}
\end{mdframed}
using the minimal coding as follows:
\begin{mdframed}[backgroundcolor=mintedbg, linecolor=mintedbg, innerleftmargin=0, innertopmargin=0,innerbottommargin=0]
\begin{python}
problem = systemID()
problem.identifying(data)
\end{python}
\end{mdframed}

The benefit of this approach is that users can easily set up  problems with minimal coding, and control them with a configuration file. A typical configuration file contains two part: controls for VSI such as ``identify\_strategy" and controls for stepwise regression such as using ridge regression, setting ``F\_criteria" and others. A representative configuration file is presented below.

\begin{mdframed}[backgroundcolor=mintedbg, linecolor=mintedbg, innerleftmargin=0, innertopmargin=0,innerbottommargin=0]
{\centering\textbf{Example of configuration file}}
\begin{python}
[VSI]
data_dir=N/A
identify_strategy= specified_target
target_index= 0

[StepwiseRegression]
basis_drop_strategy = most_inignificant
regression_method = ridge
alpha_ridge = 1.0e-5
F_criteria=1
\end{python}
\end{mdframed}
We refer readers to our previous work \cite{WangCMAME2019, WangCMAME2021,Wang2020-MRu} and online documentation for the detailed explanations.

The full script for discovering the pattern forming physics problem described in Section \ref{sec:pattern} is then presented as following:
\begin{mdframed}[backgroundcolor=mintedbg, linecolor=mintedbg, innerleftmargin=0, innertopmargin=0,innerbottommargin=0]
\begin{python}
mechanoChemML.examples.systemID.Example1_pattern_forming.main
\end{python}
\end{mdframed}

\begin{figure}
\centering
\includegraphics[width=0.8\textwidth]{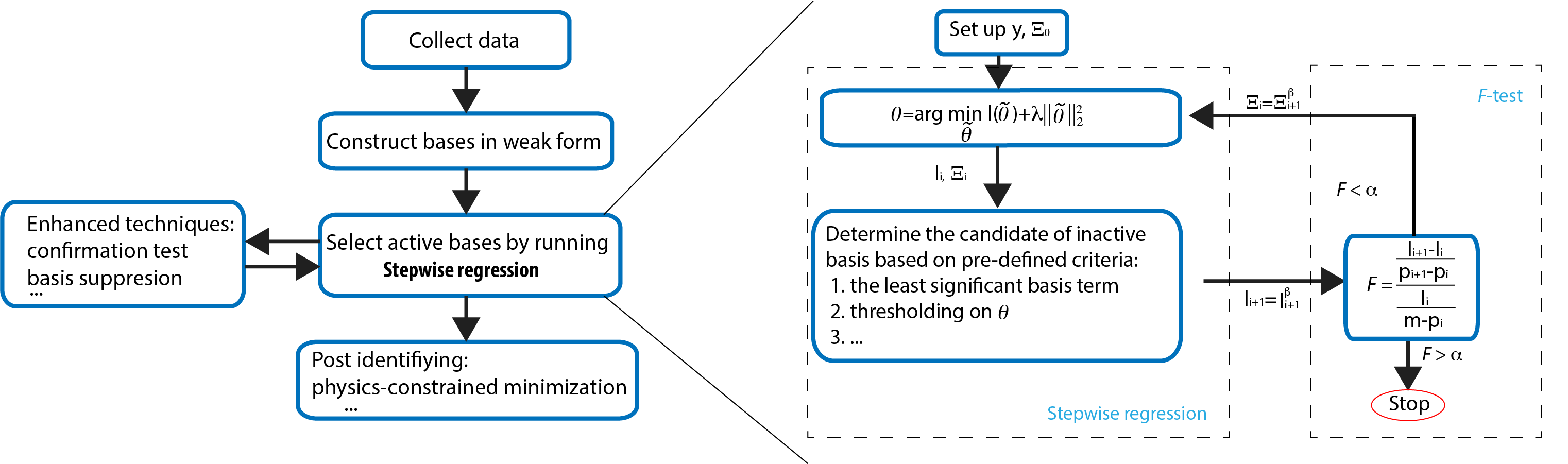}
\caption{Flowchart of Variational System Identification as a workflow.}
\label{fig:flowchart_systemID}
\end{figure}

The VSI workflow is also illustrated in Figure \ref{fig:flowchart_systemID}.

\subsubsection{Multi-resolution learning}\label{sec:workflow-homogenization}
\begin{figure}[h!]
\centering
\includegraphics[width=0.4\linewidth]{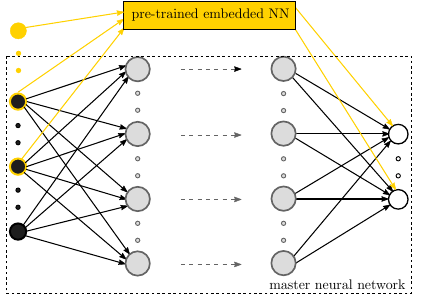}
\caption{Illustration of the NN architecture used for multi-resolution learning.}
\label{fig:multi-resolution-nn-ill}
\end{figure}

In materials physics problems, it is not uncommon to encounter data that possesses a hierarchical structure. For example, when studying the homogenized stress-strain response of a family of multi-component crystalline microstructures, the free energy of each microstructure has a multi-resolution structure with a dominant trajectory arising from phase transformations that drive evolution of the microstructure, and small-scale fluctuations from strains that explore the effective elastic response of a given microstructure \cite{Zhang2020Garikipati-CMAME-ML-RVE}. The dominant trajectory strongly depends on the microstructural information, such as the volume fraction, the location and orientation of each crystalline phase, and the interfaces, whereas the small-scale fluctuations are related to the applied loading.
Multi-resolution learning, as illustrated in Fig. \ref{fig:multi-resolution-nn-ill}, can be used to capture the details in the data, which are not well-delineated by the pre-trained model.
Taking the homogenization of microstrutures problem presented in Section \ref{sec:example-homogenization} as an example,
a multi-resolution neural network (MRNN) can be built upon pre-trained fully connected NNs or convolutional NNs, which describe the dominant part of the free energy, to learn the small-scale fluctuations of free energy and predict homogenized stresses.
The loss function for such a problem therefore is written as
\begin{equation}
\text{MSE} = \frac{1}{m} \sum_{i} \left( \mathbf{Y}- \mathbf{Z} \right)_i^2
\quad \text{with} \quad
\mathbf{Y}= \Psi_\text{mech} - \Psi_\text{mech,NN}^0
\label{eq:new-mse}
\end{equation}
where $\mathbf{Y}$ is the label, $\mathbf{Z}$ is the MRNN predicted value, $\Psi_\text{mech}$ is the hierarchical quantity to be learned, and $\Psi_\text{mech,NN}^0$ is the pre-trained NN learned dominant information in the data.
If additional physics-based constraints need to be applied, the loss function could be updated as
\begin{equation}
\text{MSE} = \frac{1}{m} \sum_{i} \left[ \left( \mathbf{Y}- \mathbf{Z} \right)_i^2
+ \beta \left\Vert \BP_\text{MRNN} -  \BP_\text{DNS} \right\Vert_i^2\right]
\quad \text{with} \quad
\mathbf{Y}= \Psi_\text{mech} - \Psi_\text{mech,NN}^0
\label{eq:new-mse-penalization}
\end{equation}
where $\beta$ is a parameter to penalize the constraints expressed in terms of the Frobenius norm $\left\Vert \bullet \right\Vert$.
Multi-resolution learning is a two-step learning process, which can be constructed based on classes and functions provided in the \texttt{mechanoChemML} library, such as
\begin{mdframed}[backgroundcolor=mintedbg, linecolor=mintedbg, innerleftmargin=0, innertopmargin=0,innerbottommargin=0]
\begin{python}
mechanoChemML.workflows.mr_learning.mrnn_models
mechanoChemML.workflows.mr_learning.mrnn_utility.
\end{python}
\end{mdframed}
See the homogenization of microstrutures problem presented in Section \ref{sec:example-homogenization} for details of constructing a MRNN.

\subsubsection{NN-based solver for PDEs}\label{sec:workflow-pde}
\begin{figure}[h!]
\centering
\includegraphics[width=1.0\textwidth]{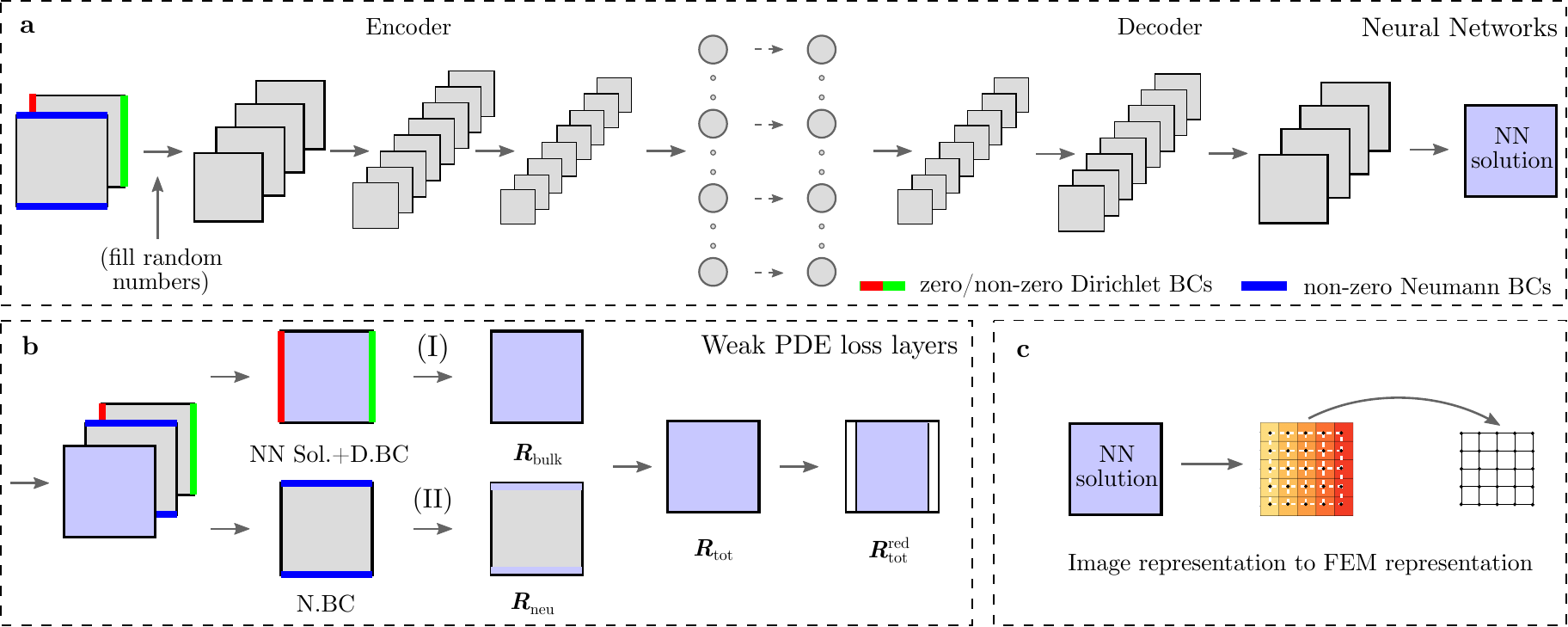}
\caption{Illustration of the NN architecture and steps in the residual calculation for a NN-based PDE solver.}
\label{fig:pde-nn-ill}
\end{figure}
We have developed NN-based PDE solvers that compute full field solutions orders of magnitude faster than the traditional PDE solvers \cite{Zhang2021Garikipati-BNN-weak-solution-PDE-SS}. These solvers work for both small datasets, which only contain a few boundary value problems (BVPs), and large datasets, which could contain hundreds of thousands of BVPs.
The workflow to construct a NN-based PDE solver consists of (i) implementing the discretized residual of PDEs (ii) building PDE constrained NNs, (iii) preparing BCs-encoded NN input data,  (iv) training NNs, and (v) testing NNs.
The NN inputs are multi-dimensional \verb numpy ~arrays, which can be easily generated.
One can use the \verb mechanoChemML.workflows.pde_solver.geometry ~ to generate BCs-encoded NN inputs for arbitrary polygons.
The PDE constrained NNs are constructed as subclasses of the \verb tf.keras.Model .
This allows us to use methods such as \verb train_on_batch ~and \verb test_on_batch ~from the Keras model to train and test the PDE constrained NNs.

To construct NNs to solve PDEs, we use \verb mechanoChemML.src.nn_models.BNN_user_weak_pde_general , whose output also contains the NN input information, including the BCs and domain, to construct the PDE constrained loss.

A templated workflow for solving general elliptic PDEs is provided at
\begin{mdframed}[backgroundcolor=mintedbg, linecolor=mintedbg, innerleftmargin=0, innertopmargin=0,innerbottommargin=0]
\begin{python}
mechanoChemML.workflows.pde_solver.pde_workflow_steady_state,
\end{python}
\end{mdframed}
which utilizes classes and functions provided at
\begin{mdframed}[backgroundcolor=mintedbg, linecolor=mintedbg, innerleftmargin=0, innertopmargin=0,innerbottommargin=0]
\begin{python}
mechanoChemML.src.pde_layers
mechanoChemML.src.nn_models,
\end{python}
\end{mdframed}
with the detailed steps to compute the discretized residual being provided at
\begin{mdframed}[backgroundcolor=mintedbg, linecolor=mintedbg, innerleftmargin=0, innertopmargin=0,innerbottommargin=0]
\begin{python}
mechanoChemML.workflows.pde_solver.pde_workflow_steady_state._compute_residual.
\end{python}
\end{mdframed}
The discretized residual can then be used to construct either the loss for deterministic NNs or the likelihood function that is used in the total loss of probabilistic NNs. Ref. \cite{Zhang2021Garikipati-BNN-weak-solution-PDE-SS} has details on the theoretical background to the construction of different loss functions.

For BNNs, the model parameters $\BTheta$ are stochastic and  sampled from a posterior distribution $P(\BTheta | \calD)$, which is computed based on Bayes' theorem
\begin{equation}
P(\BTheta | \calD) = \frac{P(\calD|\BTheta) P(\BTheta)}{P(\calD)},
\label{eq:bayes}
\end{equation}
where $\calD$ denotes the i.i.d.~observations (training data) and $P$ represents the probability density function.
In \eref{eq:bayes}, $P(\calD|\BTheta)$ is the likelihood, $P(\BTheta)$ is the prior probability, and $P(\calD)$ is the evidence, respectively.
The likelihood is the probability of the observed data $\calD$ given parameters $\BTheta$.

To compute the posterior distributions of $\BTheta$, we use variational inference, which approximates the exact posterior distribution $P(\BTheta | \calD)$ with a more tractable surrogate distribution $Q(\BTheta)$ by minimizing the Kullback-Leibler (KL) divergence
\begin{equation}
Q^* = \text{arg~min}~D_\text{KL}(Q(\BTheta)||P(\BTheta | \calD)).
\label{eq:kl-divergence}
\end{equation}
with the KL divergence being computed as
\begin{equation}
D_\text{KL}(Q(\BTheta)||P(\BTheta | \calD)) = \mathbb{E}_Q[\log Q(\BTheta)] - \mathbb{E}_Q[\log P(\BTheta, \calD)]  + \log P(\calD).
\label{eq:kl-divergence-calculation}
\end{equation}
Since the evidence $P(\calD)$ is very expensive to compute, and in general challenging to obtain, we optimize the so-called evidence lower bound (ELBO),  which is equivalent to the KL-divergence up to an added constant, with
\begin{equation}
\begin{aligned}
\text{ELBO}(Q) &= \mathbb{E}_{Q}[\log P(\calD | \BTheta )] - D_\text{KL}\left( Q(\BTheta) ||P(\BTheta)  \right).
\end{aligned}
\label{eq:elbo}
\end{equation}
The loss function for the BNN is written as
\begin{equation}
\calL = D_\text{KL}\left( Q(\BTheta) ||P(\BTheta)  \right) - \mathbb{E}_{Q}[\log P(\calD | \BTheta )],
\label{eq:bnn-loss-exact}
\end{equation}
which consists of a prior-dependent part and a data-dependent part, with the former being the KL-divergence of the surrogate posterior distribution $Q(\BTheta)$ and the prior $P(\BTheta)$, and the latter being the negative log-likelihood cost, which is related to the discretized residual of a specific PDE system.
In the \texttt{mechanoChemML} library, one can provide the detailed implementation of a specific PDE system in the method \texttt{\_bulk\_residual}.
The bulk residual for the specific PDE system can be constructed by utilizing the methods defined in the class \verb mechanoChemML.src.pde_layers.LayerBulkResidual .
An example of the use of the NN-based PDE solver for steady-state diffusion is provided in Section \ref{sec:example-steady-state-diffusion}.

\section{Scientific application examples}
\label{sec:sciexamples}

\subsection{Active learning}

\subsubsection{Learning free energy density functions for scale bridging}
\label{sec:free_energy}

The formation and evolution of material microstructures can be simulated at the continuum scale using phase field modeling. These techniques rely on a free energy density function for the material, which, among other physics, encapsulates a description of the stable phases of the materials. Atomistic models, including DFT and statistical mechanics calculations, can be used to obtain the gradient of the free energy, i.e. the chemical potential, for discrete values of chemical composition and/or other parameters describing the ordered/disordered state of the atomic structure, in terms of order parameters. Learning a free energy density function from these data that can be used in phase field models is a powerful tool in bridging scales.

Since the chemical potential training data that are produced are gradients of the free energy, we employ IDNNs, which were described in Section \ref{sec:idnn}, to train to the data. With order parameters $\eta_k$ and chemical potentials $\mu_k := \partial f/\partial \eta_k$, $k=1,\ldots,n$, the IDNN training involves the following optimization:
\begin{align}
\bsym{\hat{W}},\bsym{\hat{b}} =
\underset{\bsym{W},\bsym{b}}{\mathrm{arg\; min}}\,\sum_{k=1}^n\mathrm{MSE}\left(\frac{\partial f(\bsym{\eta};\bsym{W},\bsym{b})}{\partial X_k}\Big |_{\bsym{\hat{\eta}}_\theta},\hat{\mu}_{k_\theta}\right)
\end{align}
This results in the following representation of the free energy density, $\hat{f}$ as a function of the order parameters:
\begin{align}
\hat{f}(\bsym{\eta}) = f(\bsym{\eta};\bsym{\hat{W}},\bsym{\hat{b}})
\end{align}
where $f$ has the form of a standard, fully connected deep neural network.

Additionally, in some situations the number of order parameters may increase the dimensionality of the input space to the point where it is unfeasible to uniformly and densely sample the entire space. In these cases, it is helpful to employ active learning, as described in Section \ref{sec:active_learning}, to guide the sampling of data. An example is the Ni-Al system, described by us in previous work \cite{Teichert2020}. Ni$_{1-x}$Al$_x$ forms an ordered structure a $x=1/4$, while a disordered structure is formed for lower values of $x$. Additionally, four different translational variants of the $x=1/4$ ordering exist, which affect the formation of precipitates in the Ni-Al system. By defining the input space of the free energy using composition, $\eta_0$ and three additional order parameters, $\eta_1$, $\eta_2$, $\eta_3$, it is possible to model the order-disorder transition and track the formation of variants. The stable disordered phase at low $x$ and each of the four ordered variants at $x=1/4$ correspond to wells in the free energy density. Because of the importance is capturing these energy wells, we used pointwise error and local convexity as criteria to guide the local sampling.

\begin{figure}
\centering
\includegraphics[width=0.9\textwidth]{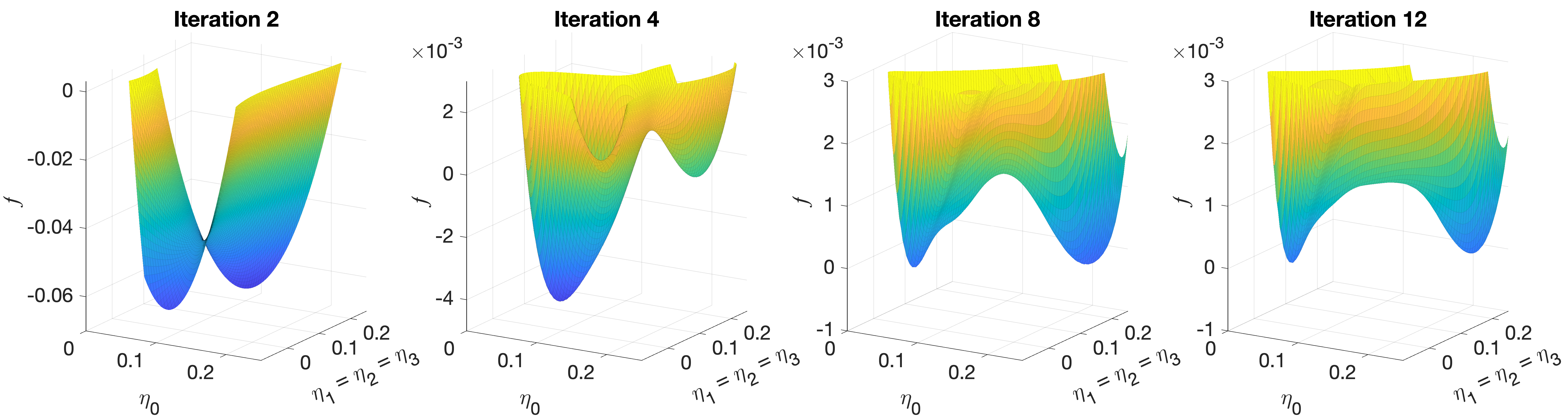}
\caption{Plots of the free energy density (i.e., the integrated IDNN) in a two-dimensional subspace at various iterations of the active learning workflow. (A linear function of $\eta_0$ has been added to each free energy surface to ``rotate'' the surface and enhance visualization of the convex regions.)}
\label{fig:free_en}
\end{figure}

As implemented in the \texttt{mechanochemML} library, the IDNN was considered sufficiently converged after twelve iterations of the active learning workflow. The evolution of a two-dimensional slice of the free energy is shown in Figure \ref{fig:free_en}, including energy wells corresponding to the disordered phase and one of the ordered variants. The following function from the \texttt{Active\_learning} class encapsulates the active learning workflow:
\begin{mdframed}[backgroundcolor=mintedbg, linecolor=mintedbg, innerleftmargin=0, innertopmargin=0,innerbottommargin=0]
\begin{python}
def main_workflow(self):
for rnd in range(self.N_rnds):
self.global_sampling(2*rnd)

if rnd==1:
self.hyperparameter_search(rnd)
custom_objects = {'Transform': Transform(self.IDNN_transforms())}

unique_inputs = self.idnn.unique_inputs
self.idnn = keras.models.load_model('idnn_1',
custom_objects=custom_objects)
self.idnn.unique_inputs = unique_inputs

self.surrogate_training(rnd)
self.local_sampling(2*rnd+1)
\end{python}
\end{mdframed}

\subsection{Multi-resolution learning}
In this section, we illustrate how to use \verb mechanoChemML ~library to perform multi-resolution learning to study the homogenized behavior of microstructures. Readers are referred to \cite{Zhang2020Garikipati-CMAME-ML-RVE} for more details.

\subsubsection{Microstructure homogenization}\label{sec:example-homogenization}
In this example, we illustrate how to use multi-resolution NNs provided in the \verb mechanoChemML ~library to construct surrogate NN models to predict the homogenized, macroscopic, mechanical free energy and stress fields arising in a family of multi-component crystalline solids that develop microstructure. The physics is driven by a non-convex free energy density function $\psi$,
\begin{equation}
\psi (c, \Be, \nabla c, \nabla\Be) = \scrF(c, \Be) + \scrG(\nabla c, \nabla \Be),
\label{eq:general-free-energy}
\end{equation}
with
\begin{subequations}\label{eq:2d-psi}
\begin{alignat}{2}
\scrF(c,\Be)
& =  \underbrace{16 d_c c^4 - 32 d_c c^3 + 16 d_c c^2}_\text{\tiny{chemical}}+ \underbrace{\frac{2d_e}{s_e^2}(e_1^2 + e_3^2) + \frac{d_e}{s_e^4}e_2^4}_\text{\tiny{mechanical}} + \underbrace{(1-2c) \frac{2d_e}{s_e^2}e_2^2}_\text{\tiny{mechanochemical}} \label{eq:2d-F}\\
\scrG(\nabla c, \nabla \Be)
& = \underbrace{\frac{1}{2} \nabla c \cdot \kappa \nabla c}_\text{\tiny{chemical}} + \underbrace{\frac{1}{2} \nabla e_2 \cdot \lambda_e \nabla e_2}_\text{\tiny{mechanical}} \label{eq:2d-G}
\end{alignat}
\end{subequations}
where $c$ is the compositional parameter, $\Be$ is the mechanical strain vector with $e_1$, $e_2$, and $e_3$ as its components, and $\{d_c$, $d_e$, $s_e$, $\kappa$, $\lambda_e\}$ are material parameters.
Here, $\scrF$ represents a homogeneous contribution from both composition and strain, and $\scrG$ being a gradient-dependent, inhomogeneous contribution to regularize the free energy density.
The resulting microstructure, driven by this free energy with its perturbation, has a hierarchical--or multi-resolution--structure, which can be resolved by an MRNN.
The main steps to build the surrogate homogenization model workflow include: (i) synthetic training data generation, (ii) training NNs to describe the dominant characteristic of the data with a hyper-parameter search, (iii) constructing the MRNN to fully represent the hierarchically evolving free energy with a hyper-parameter search.
To setup the workflow, we use the following classes in the \verb mechanoChemML ~library
\begin{mdframed}[backgroundcolor=mintedbg, linecolor=mintedbg, innerleftmargin=0, innertopmargin=0,innerbottommargin=0]
\begin{python}
mechanoChemML.src.kfold_train
mechanoChemML.src.hparameters_cnn_grid
mechanoChemML.src.hparameters_dnn_grid
mechanoChemML.workflows.mr_learning.mrnn_models
mechanoChemML.workflows.mr_learning.mrnn_utility
mechanoChemML.third_party.dns_wrapper.dns_wrapper
\end{python}
\end{mdframed}

\begin{figure}[t!]
\centering
\subfigure[DNN predicted $\Psi_\text{mech}^0$]         {\includegraphics[width=0.24\linewidth]{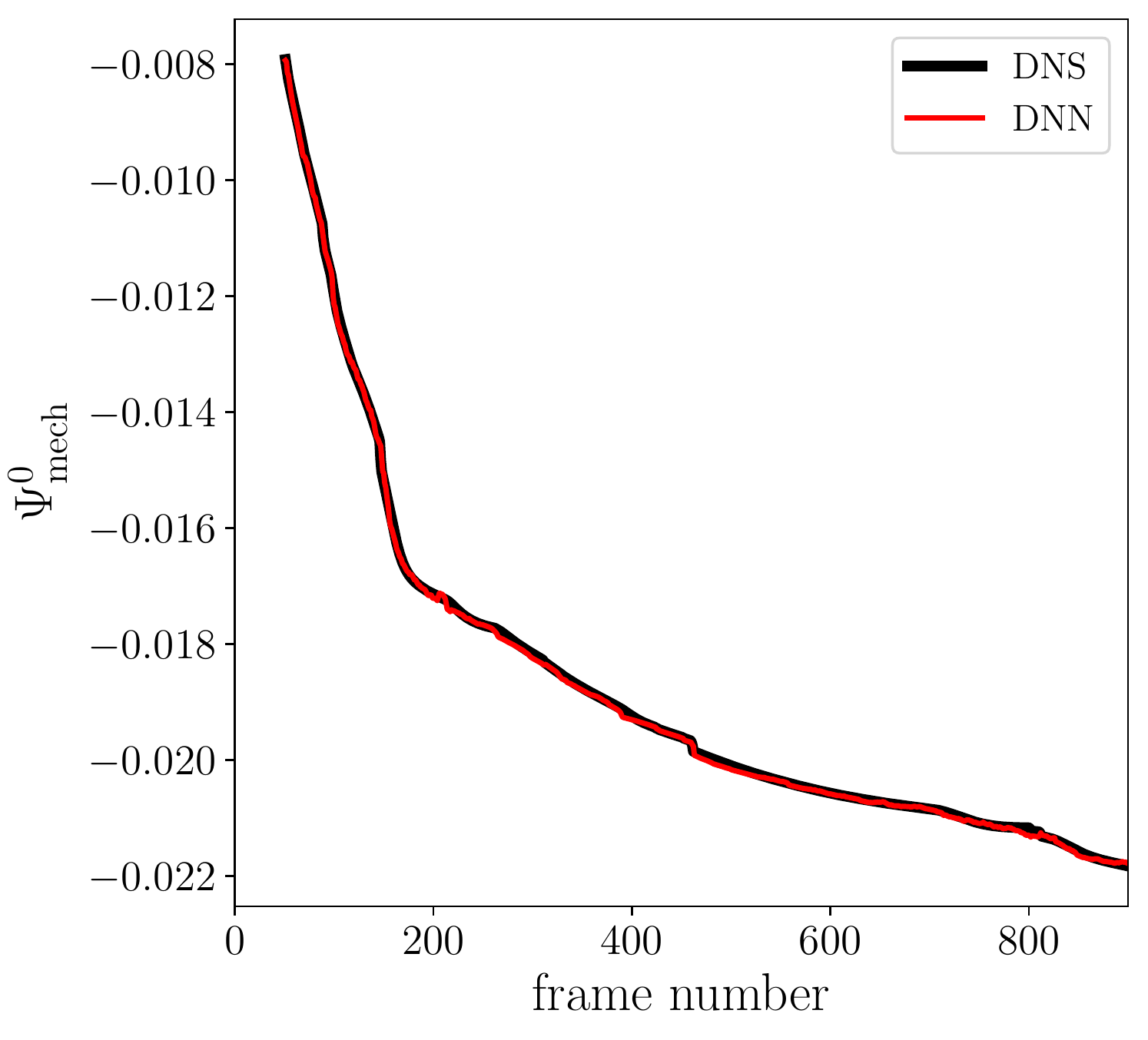}}
\subfigure[MRNN predicted  $\Delta \Psi_\text{mech}$]  {\includegraphics[width=0.26\linewidth]{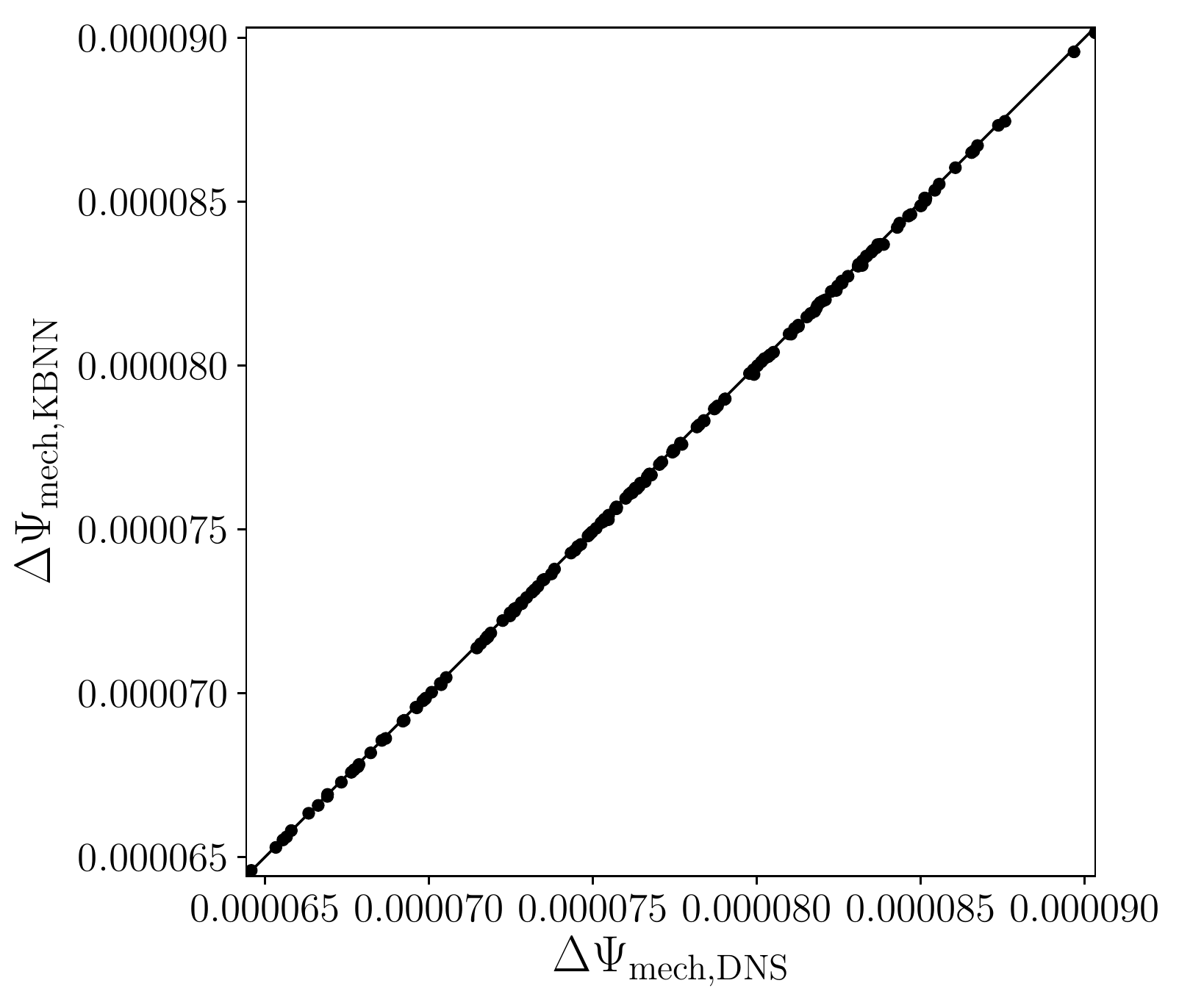}}
\subfigure[MRNN predicted $P_{11}$]{\includegraphics[width=0.24\linewidth]{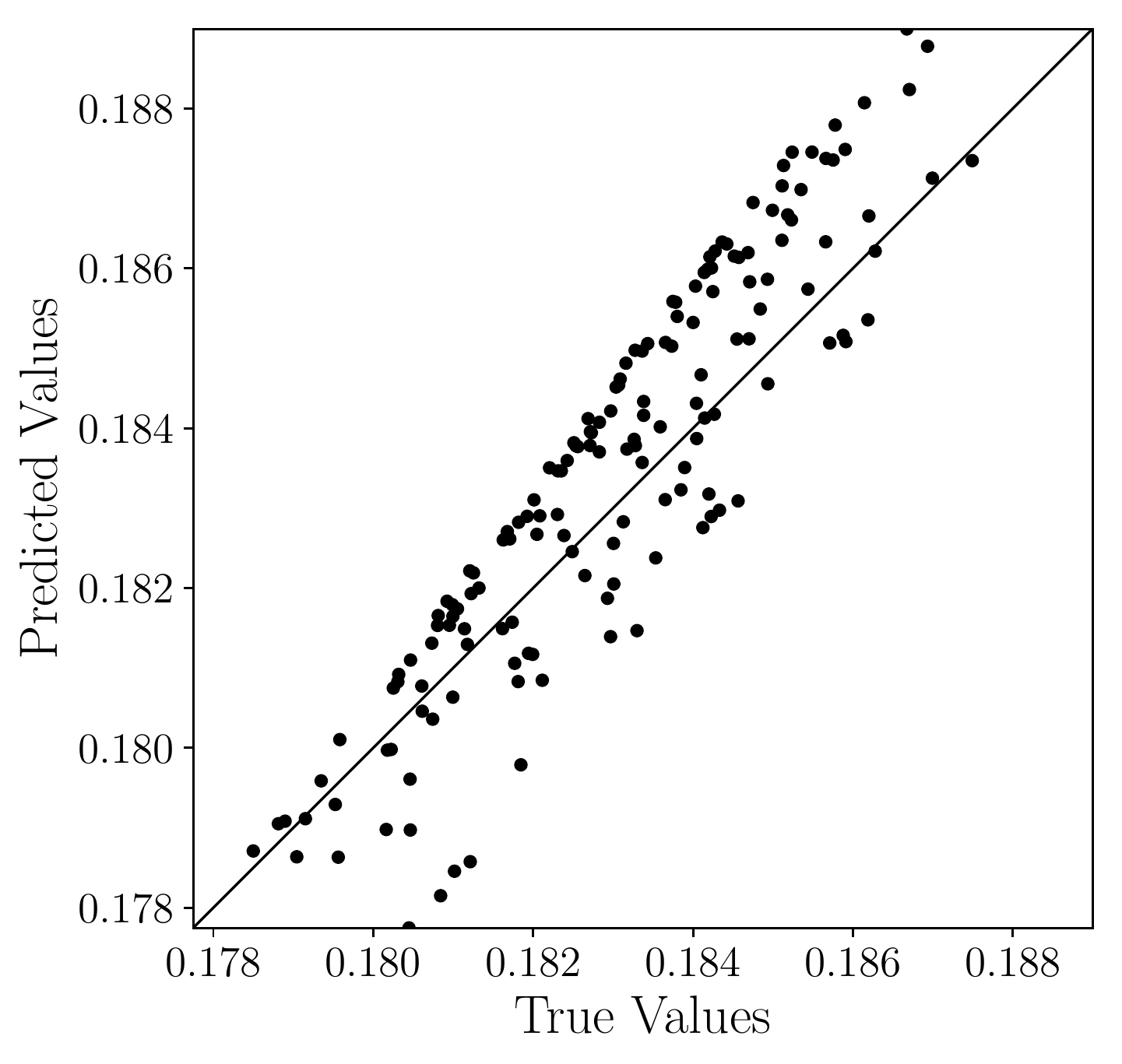}}
\subfigure[MRNN predicted $P_{22}$]{\includegraphics[width=0.24\linewidth]{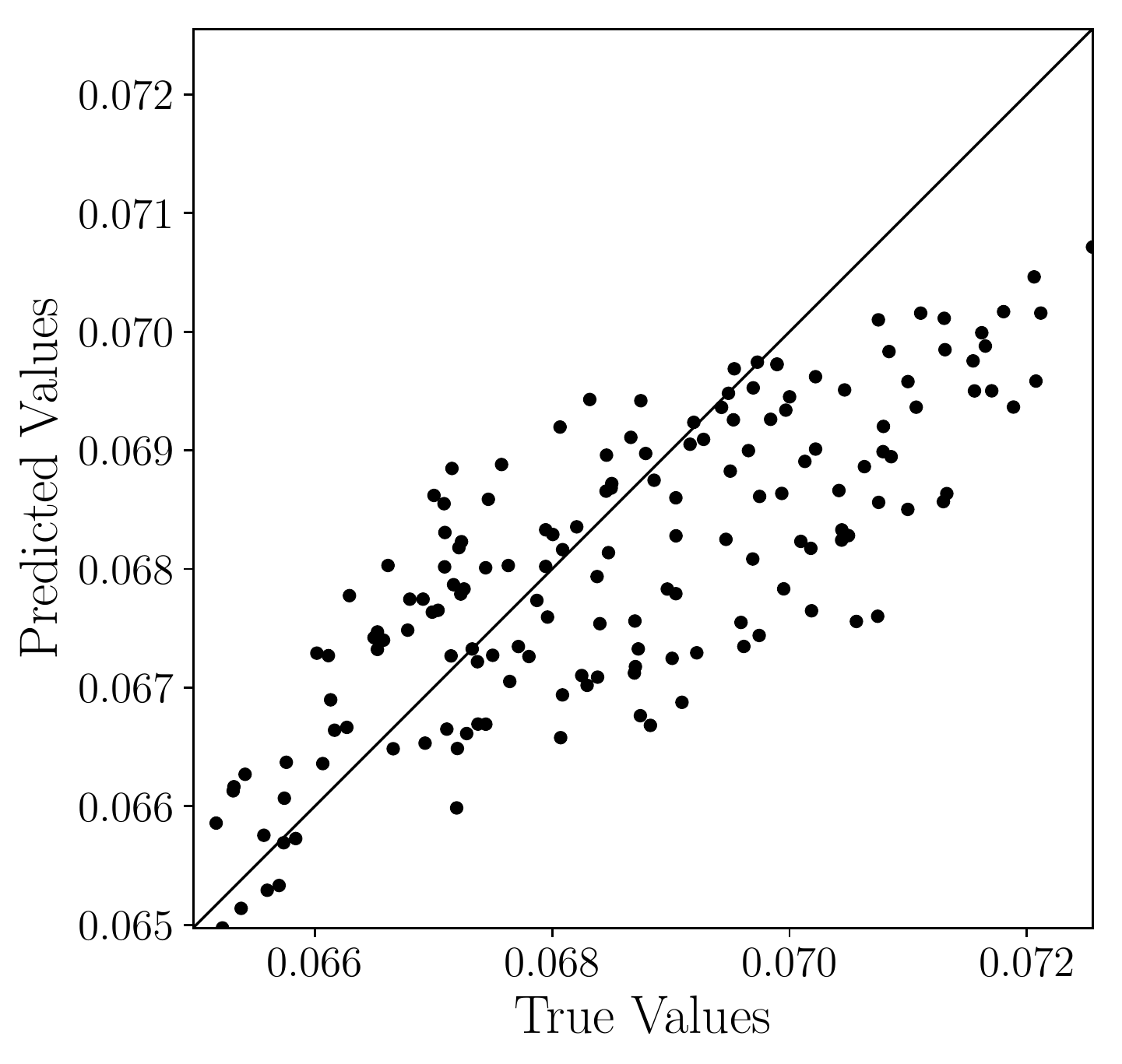}}
\caption{Illustration of the NN predicted solutions.
(a) NN predicted base mechanical free energy $\Psi_\text{mech}^0$ vs. DNS solutions.
(b) the MRNN predicted $\Delta \Psi_\text{mech}$ vs. DNS solutions.
(c-d) the components of $\BP_\text{MRNN}$ vs. DNS solutions, where the MRNN learns a reasonable derivative representation for $P_{11}$ and $P_{22}$.}
\label{fig:homogenization-results}
\end{figure}

Synthetic data generation is a tedious process and is not the focus of the \texttt{mechanoChemMl} library, although a user could potentially use the example Python DNS interface class provided at \verb mechanoChemML.third_party.dns_wrapper.dns_wrapper ~to run, the spinodal-decomposition initial boundary value problem (IBVP) that generated the data for the example in this section.
A small fraction of preprocessed data that was thus generated and used in Ref. \cite{Zhang2020Garikipati-CMAME-ML-RVE} is provided with this example in \verb mechanoChemML ~to allow readers to skip the data generation process and directly learn the workflow.

As discussed in \cite{Zhang2020Garikipati-CMAME-ML-RVE}, both DNNs and CNNs can be used to learn the dominant characteristic of the data.
In this example, we focus on DNNs in the interest of simplicity.
Furthermore, we only focus on predicting the free energy for microstructures generated from one DNS, and the homogenized stress for a single microstructure.
However, more complex examples studied in \cite{Zhang2020Garikipati-CMAME-ML-RVE} are also included in the \verb mechanoChemML ~library. Readers may refer to the code documentation for more details.
The MRNN constructed based on DNNs for microstructures from one DNS resides at
\begin{mdframed}[backgroundcolor=mintedbg, linecolor=mintedbg, innerleftmargin=0, innertopmargin=0,innerbottommargin=0]
\begin{python}
examples/mr_learning/Example1_single_microstructure_dnn
\end{python}
\end{mdframed}
with four steps
\begin{mdframed}[backgroundcolor=mintedbg, linecolor=mintedbg, innerleftmargin=0, innertopmargin=0,innerbottommargin=0]
\begin{python}
step1_hp_search_main
step2_final_dnn_main
step3_hp_search_mrnn_detail
step4_final_mrnn_no_penalize_P.
\end{python}
\end{mdframed}
The first two steps perform a hyper-parameter search to identify the best DNN structure and subsequently train the best model to capture the dominant characteristic of the data.
The MRNN is constructed by subtracting the dominant characteristic predicted by the pre-trained DNNs from the data.
The third step performs another hyper-parameter search to identify the best MRNN structure to capture the detailed characteristic, and the fourth step trains the best found model. The resulting solutions from both the DNN and MRNN are presented in Fig. \ref{fig:homogenization-results}. See Ref. \cite{Zhang2020Garikipati-CMAME-ML-RVE} for detailed interpretation of the results.

\subsection{NN-based PDE solver}

In this section, we provide an example to illustrate the use of the \verb mechanoChemML ~library to solve PDEs with NNs. See Ref. \cite{Zhang2021Garikipati-BNN-weak-solution-PDE-SS} for further details.

\subsubsection{Solving steady-state diffusion}\label{sec:example-steady-state-diffusion}

\begin{figure}[t!]
\centering
\subfigure[setup]{\includegraphics[width=0.17\linewidth]{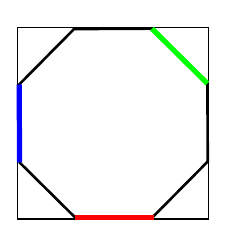}}
\subfigure[results ($32\times32$)]{\includegraphics[width=0.75\linewidth]{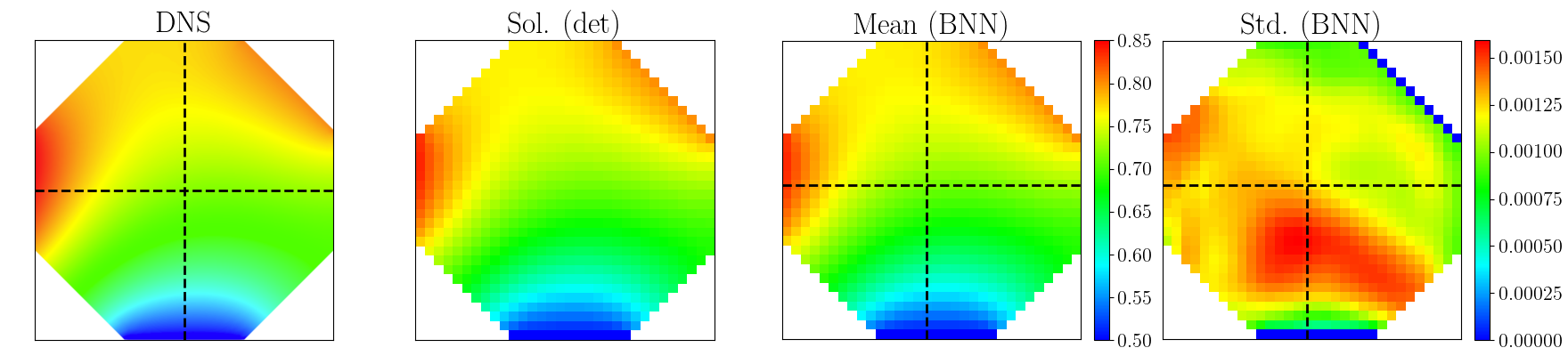}} \\
\subfigure[UQ horizontal ($32\times 32$)]{\includegraphics[trim=0.5cm 0.0cm 0.5cm 0.0cm, clip, width=0.32\linewidth]{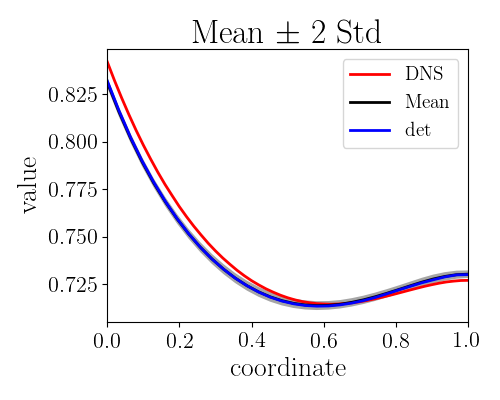}}
\subfigure[UQ vertical ($32\times 32$)]{\includegraphics[trim=0.5cm 0.0cm 0.5cm 0.0cm, clip, width=0.32\linewidth]{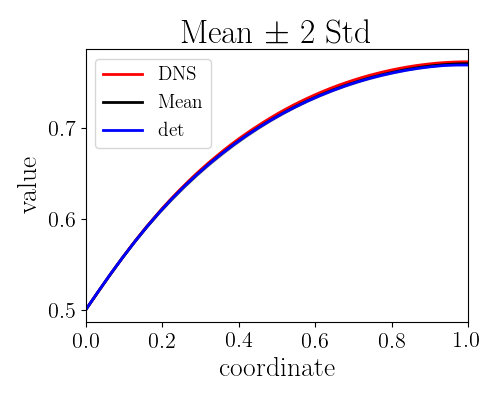}}
\caption{Steady-state diffusion BVP on an octagonal domain with mixed BCs.
(a) Simulation setup, red: zero Dirichlet BC, green: non-zero Dirichlet BC, blue: non-zero Neumann BC.
(b) Solutions from DNS, deterministic (det) NNs, and BNNs (Mean, Std.) for an output resolution of $32\times 32$.
(c-d) Quantitative comparison of the solution distribution between DNS and BNNs along the horizontal and vertical dashed lines in (b).}
\label{fig:diffusion-irregular-1bvp}
\end{figure}

Here, we construct the NN-based solver to solve single species steady-state diffusion, whose strong form is written as
\begin{equation}
\begin{aligned}
\nabla \cdot \BH = \Bzero \quad & \text{on} \quad \Omega, \\
c (\BX) = \bar{c} (\BX)  \quad & \text{on} \quad  \Gamma^{c}, \\
H = \bar{H} (\BX)  \quad & \text{on} \quad  \Gamma^H,
\end{aligned}
\label{eq:general-pde-diffusion}
\end{equation}
In \eref{eq:general-pde-diffusion}, $c$ represents the compositional order parameter, $\BH$ is the diffusive flux term defined as
\begin{equation}
\BH = - D \nabla c,
\label{eq:diffusion-flux}
\end{equation}
with $D$ as the diffusivity, and $H$ is the outward surface flux in the normal direction.
The discretized residual function  for steady-state diffusion is written as
\begin{equation}
\begin{aligned}
\BR  & = \sum_{e=1}^{n_\text{elem}} \left\{ \int_{\Omega^e} \BB^T \BH dV - \int_{\Gamma^{e,H}} \BN^T \bar{H}~dS \right\}. \\
\end{aligned}
\label{eq:discretized-residual-diffusion}
\end{equation}
We consider a diffusivity of $D=1.0$.
The detailed Python implementation of this discretized residual of this PDE system is provided in the tutorial example in the library documentation.
We use the NN-based PDE solver to solve a boundary value problem (BVP) on an octagonal domain with mixed applied BCs. The BVP setup and the associated results are illustrated in Fig. \ref{fig:diffusion-irregular-1bvp}.
The details of this example, including the uncertainty quantification, can be found in Ref. \cite{Zhang2021Garikipati-BNN-weak-solution-PDE-SS}.

In the \verb mechanoChemML ~library, we have provided implementations of steady-state diffusion, linear elasticity, and nonlinear elasticity at
\begin{mdframed}[backgroundcolor=mintedbg, linecolor=mintedbg, innerleftmargin=0, innertopmargin=0,innerbottommargin=0]
\begin{python}
mechanoChemML.workflows.pde_solver.pde_system_diffusion_steady_state
mechanoChemML.workflows.pde_solver.pde_system_elasticity_linear
mechanoChemML.workflows.pde_solver.pde_system_elasticity_nonlinear
\end{python}
\end{mdframed}
For the steady-state diffusion example, input data and configuration files are provided at
\begin{mdframed}[backgroundcolor=mintedbg, linecolor=mintedbg, innerleftmargin=0, innertopmargin=0,innerbottommargin=0]
\begin{python}
examples/pde_solver/Example1_diffusion_steady_state
\end{python}
\end{mdframed}
To generate DNS solutions for comparison, users can follow an example Python DNS interface class provided at \verb mechanoChemML.third_party.dns_wrapper.dns_wrapper ~to run the steady-sate diffusion boundary value problem (BVP) by specifying the corresponding BVP name via the command line as:
\begin{mdframed}[backgroundcolor=mintedbg, linecolor=mintedbg, innerleftmargin=0, innertopmargin=0,innerbottommargin=0]
\begin{python}
python mechanoChemML/third_party/dns_wrapper/dns_wrapper.py -e steady_state_diffusion
\end{python}
\end{mdframed}
We note that the Python DNS wrapper only provides an interface to interact with different physics-based simulation tools. Users need to install the actual software in order to run the physics-based simulation. Furthermore, the physics-based simulation tools need to be compiled to a Python dynamic library. In this example, the installation of \verb mechanoChemFEM ~is needed. The results from the NN-based PDE solver are presented in Fig. \ref{fig:diffusion-irregular-1bvp}.

\subsection{System identification}
In this section, we demonstrate system inference in the \verb mechanoChemML ~library with an example of pattern formation in material microstructures using stepwise regression introduced in Section \ref{sec:stepwise} within the VSI framework.

\subsubsection{Pattern formation of material microstructures}
\label{sec:pattern}
\begin{figure}
\centering
\subfigure[$c_1$]{\includegraphics[width=0.3\textwidth]{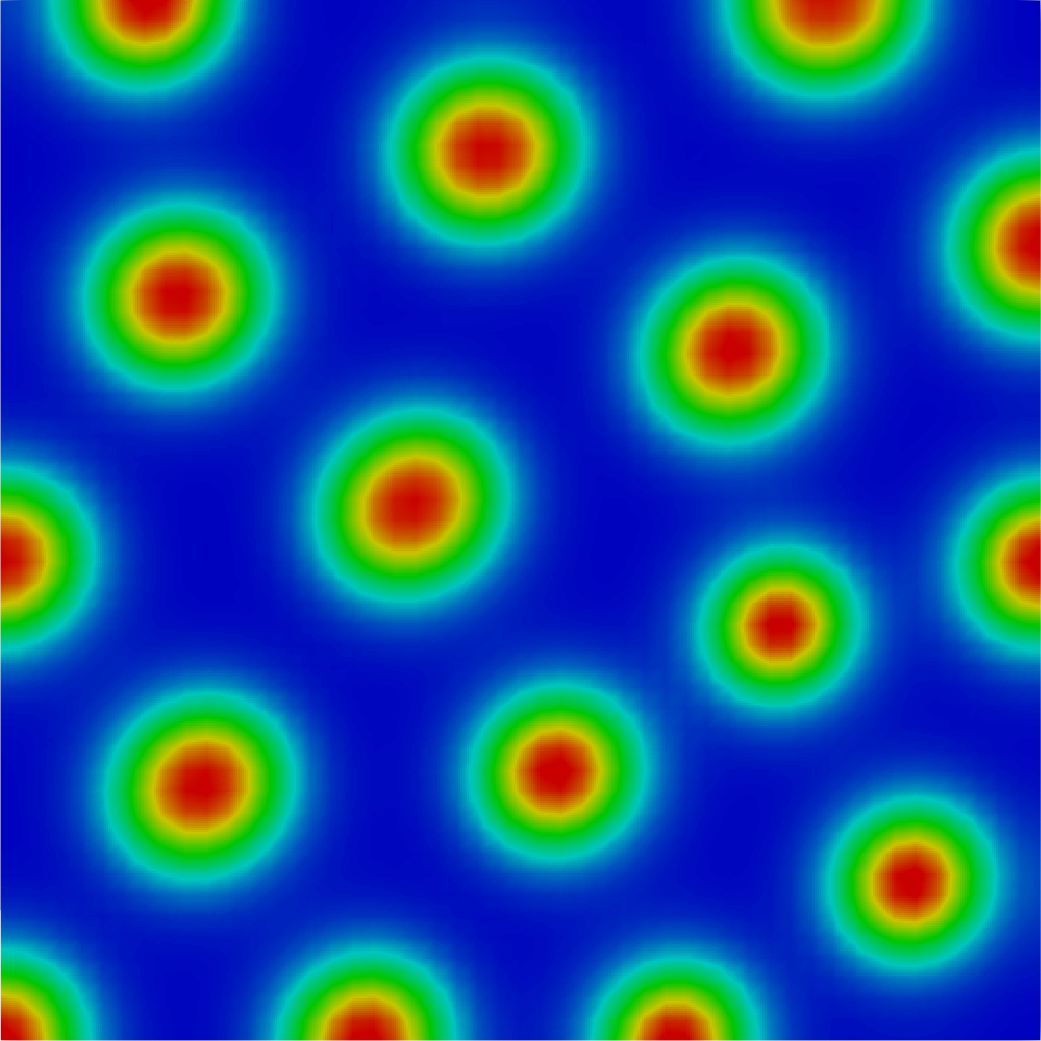}}\qquad
\subfigure[$c_2$]{\includegraphics[width=0.3\textwidth]{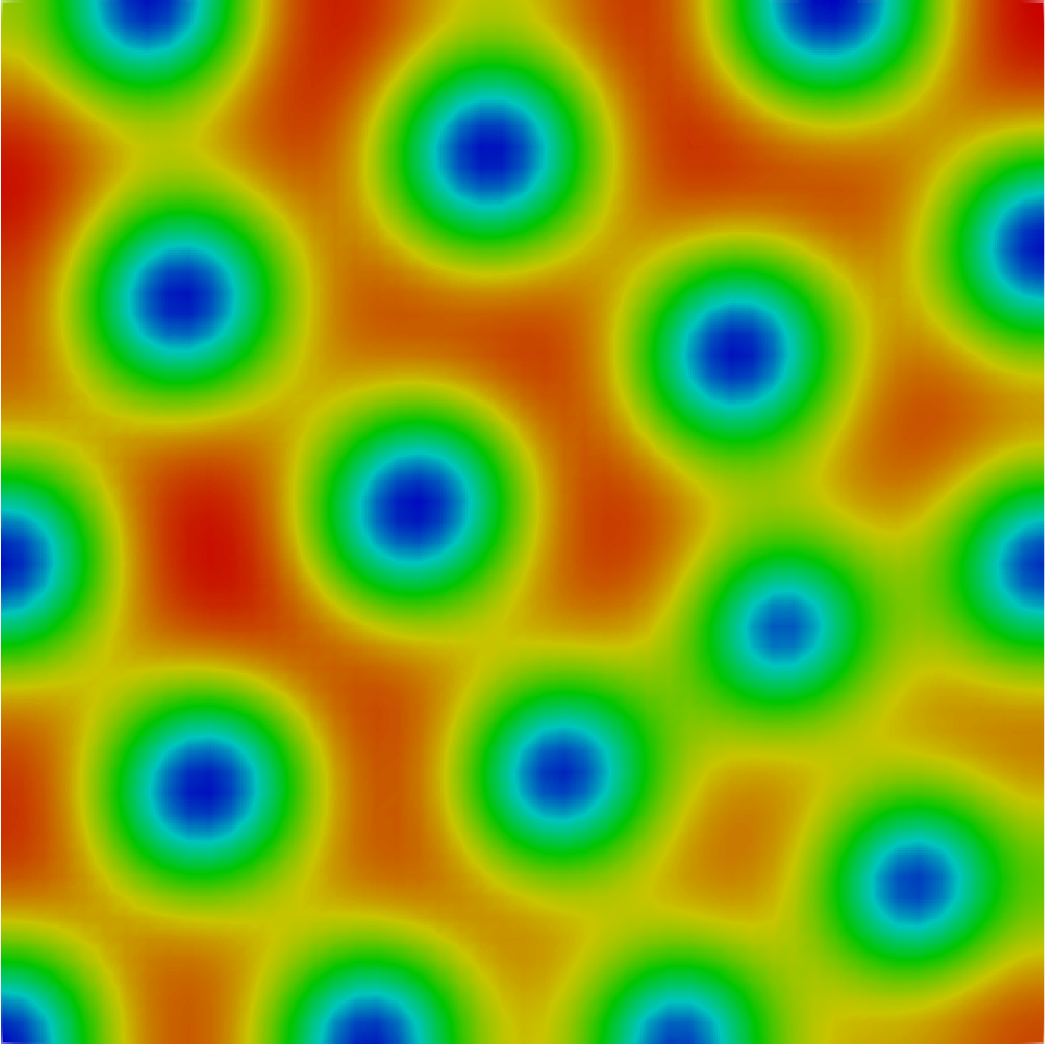}}
\caption{$c_1$ and $c_2$ at steady state, simulated using the pre-defined Schnakenberg kinetics model.}
\label{fig:patterns}
\end{figure}

In the case of dynamic pattern formation in materials, the physics is governed by first-order (in time) PDEs--a class that includes the time-dependent reaction-diffusion and phase field equations.
For demonstration, consider the following model form in $[0,T]\times\Omega$:
\begin{align}
&\frac{\partial c_1}{\partial t}=D_{11}\nabla^2c_1+D_{12}\nabla^2c_2+R_{10}+R_{11}c_1+R_{12}c_2+R_{13}c_1^2c_2
\label{eq:model_eq1}\\
&\frac{\partial c_2}{\partial t}=D_{21}\nabla^2c_1+D_{22}\nabla^2c_2+R_{20}+
R_{21}c_1+R_{22}c_2+
R_{23}c_1^2c_2\label{eq:model_eq2}\\
&\text{with} \quad \nabla c_1\cdot\mathbf{n}=0,\quad \nabla c_2\cdot\mathbf{n}=0 \text{ on }\Gamma = \partial\Omega\\
&\text{and}\; c_1\bx,0)=c_{1_0}(\bx),\quad c_2(\bx,0)=c_{2_0}(\bx).
\end{align}
Here, $c_1(\bx,t)$ and $c_2(\bx,t)$ are the compositions,  with diffusivities $D_{11},\ldots,D_{22}$ and reaction rates $R_{10},\ldots,R_{23}$ assumed constant in space and time. This model represents the coupled diffusion-reaction equations for two species following Schnakenberg kinetics \cite{Schnakenberg1976}. For an activator-inhibitor species pair having auto-inhibition with cross-activation of a short range species, and auto-activation with cross-inhibition of a long range species these equations form so-called Turing patterns \cite{Turing1952}. See Fig. \ref{fig:patterns}.

For infinite-dimensional problems with Dirichlet boundary conditions on $\Gamma^c$, the weak form corresponding to the strong form in Equation \eqref{eq:model_eq1} or \eqref{eq:model_eq2} is, $\forall \,w \in \mathscr{V}= \{w\vert ~w = ~0 \;\mathrm{on}\;  \Gamma^c\}$,
find $c$ such that
\begin{align}
\int_{\Omega}w\frac{\partial c}{\partial t}\text{d}v&=\Bomega\cdot\Bchi
\label{eq:model_weak}
\end{align}
where $\Omega$ is the domain, $\Bchi$ is the vector containing all possible independent operators in weak form:
\begin{align}
\footnotesize
\Bchi=\left[-\int_{\Omega}\nabla w\cdot\nabla c_1\text{d}v,-\int_{\Omega}\nabla w\cdot\nabla c_2\text{d}v,\int_{\Omega}w\text{d}v,\int_{\Omega}wc_1\text{d}v,\int_{\Omega}wc_2\text{d}v,\int_{\Omega}wc_1^2c_2\text{d}v\right]
\label{eq:basis_weak}
\end{align}
and $\Bomega$ is the vector of operator prefactors. Using this notation, $\Bomega=[D_{11},\ldots,R_{13}]$ for Equation (\ref{eq:model_eq1}) and $\Bomega=[D_{21},\ldots,R_{23}]$ for Equation (\ref{eq:model_eq2}).
Upon integration by parts, application of appropriate boundary conditions, and accounting for the arbitrariness of $w$, the finite-dimensionality leads to a vector system of residual equations: $\mathscr{R}=\by-\Bchi\Bomega$,

where $\by$ is the time derivative term and may  be represented via a backward difference approximation
\begin{align}
y^i=\int_{\Omega} N^i \sum_{a=1}^{n_\mathrm{b}} \frac{c_{n}^{a}-c_{n-1}^{a}}{\Delta t} N^a\text{d}v  \label{eq:basis_Ct}
\end{align}
with $N^i$ denoting the basis function corresponding to degree of freedom (DOF) $i$, and $\Delta t = t_n-t_{n-1}$ the time step. The other operators in $\Bchi$ are constructed similarly and grouped together into the matrix $\Bchi$.
Minimizing the residual norm towards $||\mathscr{R}|| = 0$ then yields the linear regression problem
\begin{align}
\by=\Bchi\Bomega.
\label{eq:least-square}
\end{align}

Solving Equation (\ref{eq:least-square}) via standard regression, especially with noisy data, will lead to a non-sparse $\Bomega$. Such a result will not sharply delineate the relevant bases for parsimonious identification of the governing system. We therefore use stepwise regression coupled with the statistical $F$-test for parsimonious inference of a minimal set of operators \cite{WangCMAME2019, WangCMAME2021}. Performing stepwise regression with the $F$-test also is a key step in the workflow of system identification (Fig. \ref{fig:flowchart_systemID}).

In another scenario, steady state, or near-steady state data with high spatial resolution may be obtained from modern microscopy methods. 
In fact the data satisfying the steady state equation:
\begin{align}
\Bchi\cdot\Btheta=0,
\label{eq:steady-state_strongForm}
\end{align}
already provide rich information about the spatial operators (that is, other than time derivatives) in the system. However, in the absence of  prior knowledge about the system, it may be challenging to choose a proper "target" operator, e.g. the left hand side of Equation (\ref{eq:least-square}). The confirmation test developed in \cite{WangCMAME2021} can provide a sharp condition for acceptance of the inferred operators. The confirmation test also has been used in the example of pattern formation in material microstructures demonstrated in this section.

\subsection{Graph-theoretic system identification}
In this section, we provide an example to illustrate the use of the graph theory based non-local calculus in the \verb mechanoChemML ~library to generate operators for Variational System Identification problems. Specifically, a 1D phase separation problem is introduced via the Allen-Cahn equations for generation of high-dimensional data. These data are used to train a reduced order model. The readers are directed to Ref. \cite{duschenes2021reduced} for an in-depth discussion on this example.

\subsubsection{Allen-Cahn dynamics}
Consider a field $\GTfield = \GTfield(x,t): \Omega \times [0,T] \mapsto \mathbb{R}$, governed by first order dynamics driven by gradient flow:
\begin{align}
\frac{\partial \GTfield }{\partial t} =&~ -M_{\GTfield}\frac{\delta \GTFreeEnergy}{ \delta \GTfield}, \qquad \text{ in } \Omega \times [0,T] \label{eq:ac_gf}\\
\nabla\GTfield\cdot \mathbf{n} =&~  0, \qquad \text{ on } \partial\Omega\\
\GTfield(x,0) = &~ \GTfield_0(x)
\end{align}
with the free energy density, $\GTFreeEnergy$ including $f$, an algebraic Landau energy density of the form
\begin{align}
\GTFreeEnergy =&~ f(\GTfield) + \frac{\lambda}{2} \vert{\nabla\GTfield}\vert^2, \qquad f(\GTfield) = (\GTfield^2 - 1)^2 \label{eq:ac_landau}
\end{align}
and $f$ having wells at $\phi = \pm 1$. The gradient energy $\lambda\vert\nabla\GTfield\vert^2$, with $\lambda > 0$ penalizes sharp transitions between the positive and negative phases $\Omega_\pm \subset \Omega$, which are defined by
\begin{equation}
x \in
\begin{cases}
\Omega_+ & \text{if }\phi(x) \ge 0\\
\Omega_- & \text{if }\phi(x) < 0
\end{cases}
\end{equation}

\begin{figure}[hpt]
\centering
\subfigure[Initial condition.] {\includegraphics[width=0.33\textwidth]{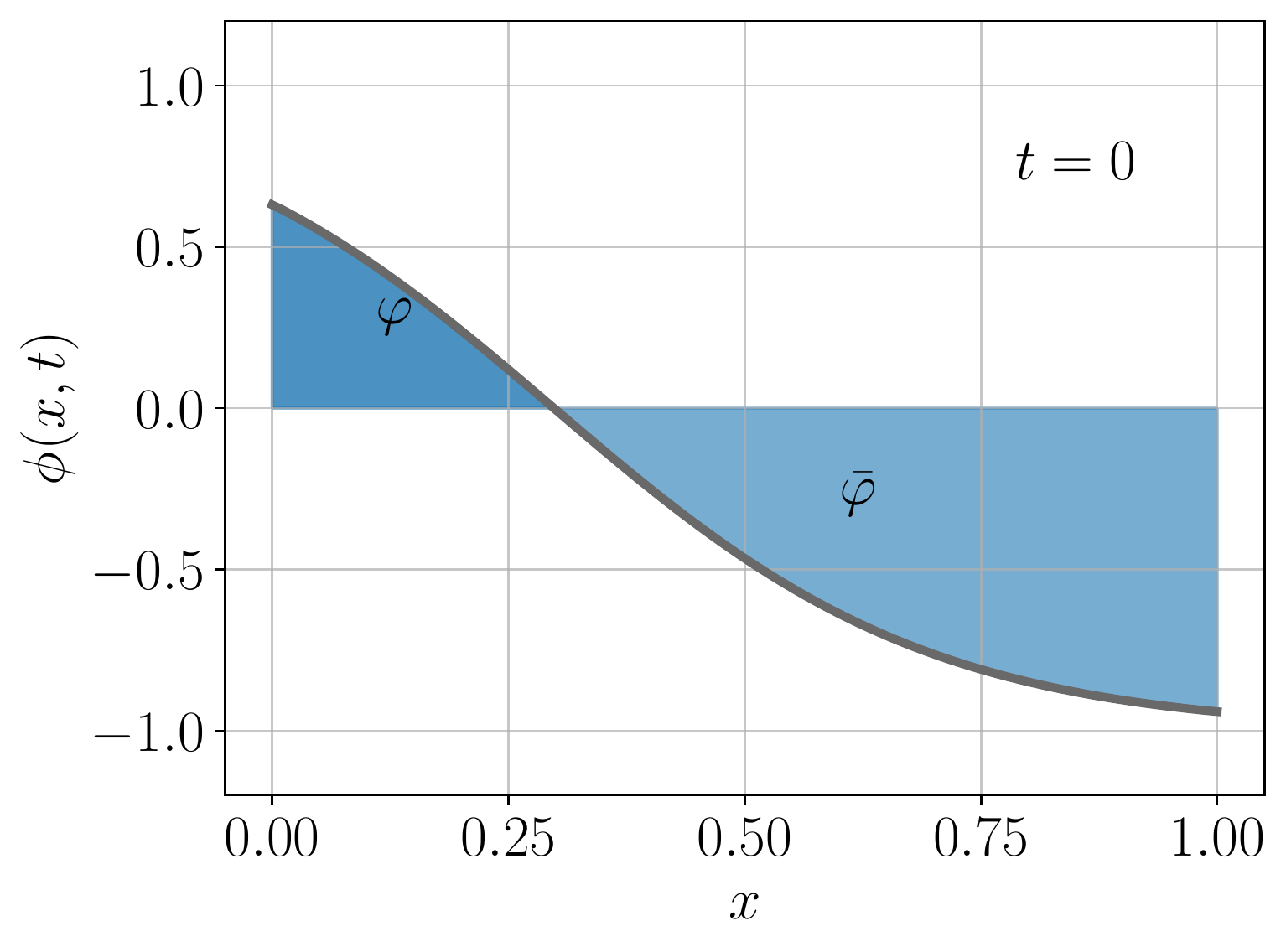}}
\subfigure[Solution at intermediate $t$.] {\includegraphics[width=0.33\textwidth]{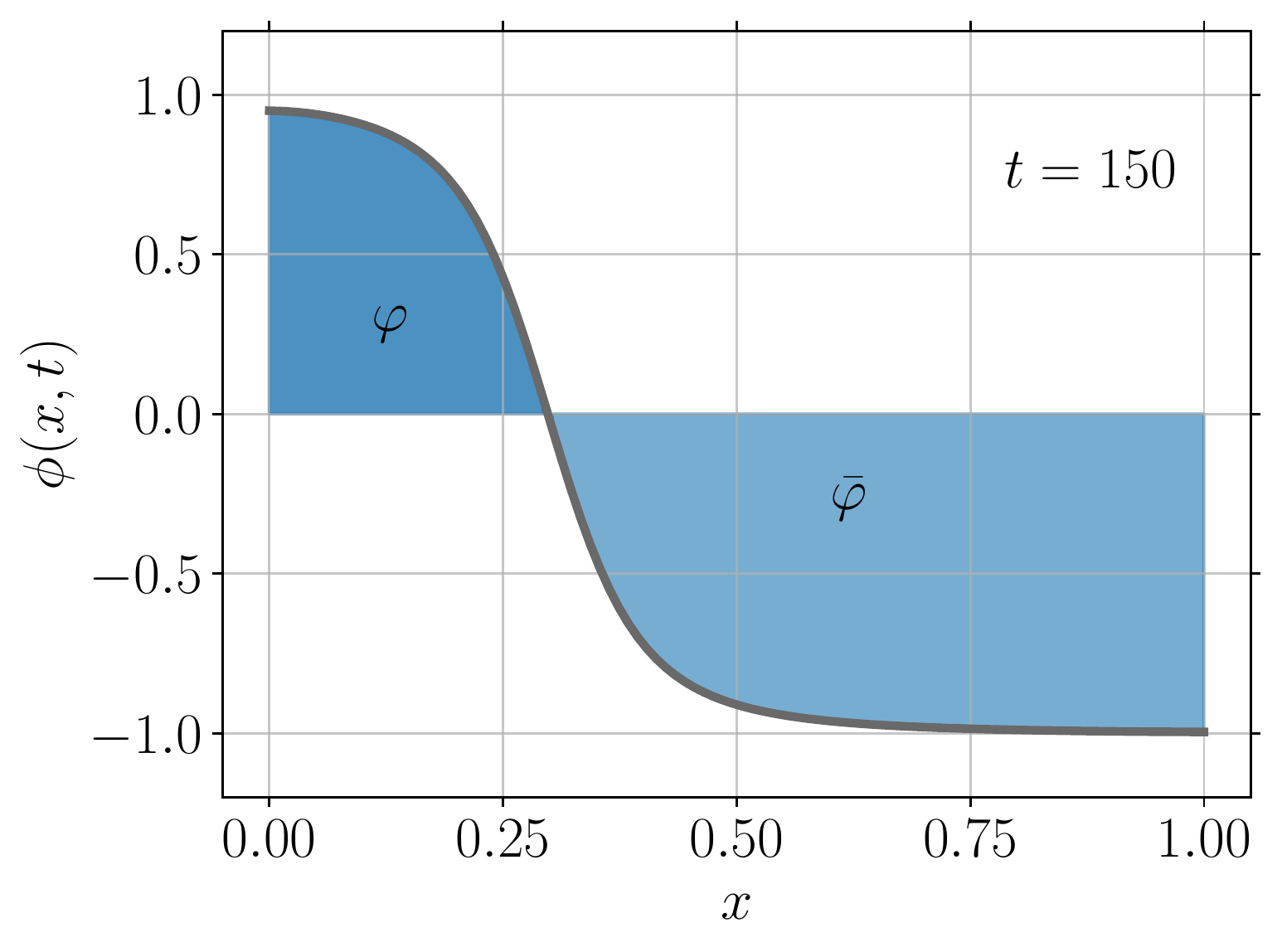}}
\subfigure[Near equilibrium solution.]
{\includegraphics[width=0.33\textwidth]{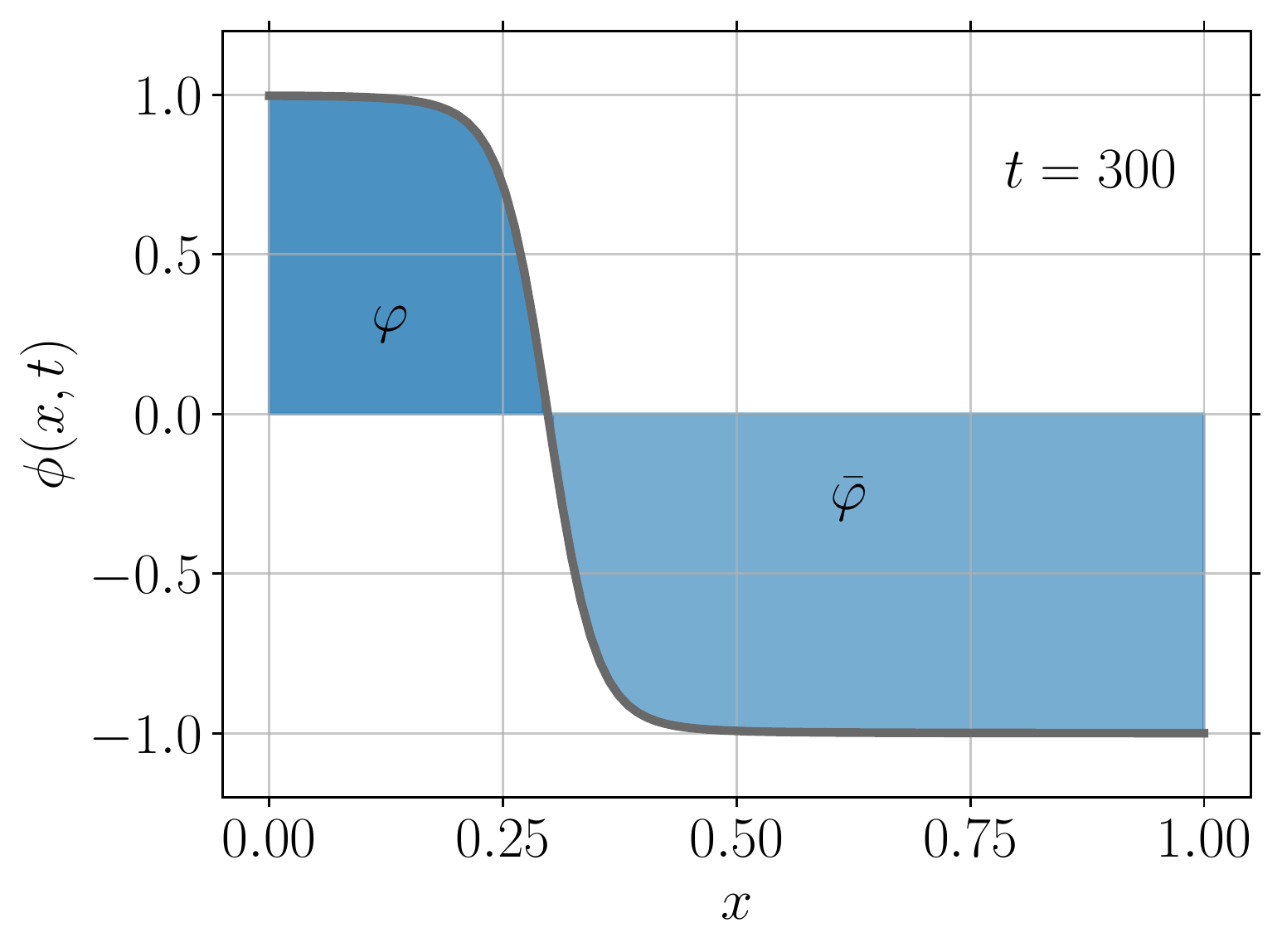}}
\caption{Field evolution of 1D Allen-Cahn dynamics with $M_{\phi} = 1e-3, \lambda = 1 $ at $0$, $150$, and $300$ time steps. A Backward-Euler scheme is used with a time step of $\Delta t = 0.01$}.
\label{fig:AllenCahn1D}
\end{figure}

The kinetics are controlled by the local mobility, $M_{\phi} \ge 0$. The Eqs \eqref{eq:ac_gf} and \eqref{eq:ac_landau} together constitute the Allen-Cahn equation \cite{Allen1979}. An example of the system dynamics in 1D appears in Fig~\ref{fig:AllenCahn1D}.

The local variations in the field, $\GTfield$, can be studied in terms of global observables, for instance, the total energy of the system, $\ObsFreeEnergy$:
\begin{align}
\ObsFreeEnergy\left[ \GTfield \right] =&~ \int\limits_{\Omega} \GTFreeEnergy(x,t)\mathrm{d}\Omega
\end{align}
A rich set of global observables can be constructed by computing phase average quantities as different functions of $\GTfield$ as:
\begin{align*}
\varphi_{g(\GTfield)_\pm} = \frac{1}{|\Omega|} \int\limits_{\Omega} g( \GTfield) I(\pm\GTfield) \mathrm{d}\Omega, \qquad I(\GTfield) = \begin{cases}
1 & \text{if }\GTfield \ge 0\\
0 & \text{if }\GTfield <0,
\end{cases}
\end{align*}
The following functions, $g(\phi)$ are chosen in this study:
\begin{align*}
g(\GTfield) \in \mathcal{G} =  \left\lbrace  \GTfield, \GTfield^2, \GTfield^3, \GTfield^4, \GTfield^5 , f(\GTfield), f'(\GTfield),\Delta \GTfield, \vert \nabla \phi \vert^2  \right\rbrace
\end{align*}
Similarly, the total energy of the system in the positive phase, $\ObsFreeEnergy_+$ can be estimated as:
\begin{align}
\ObsFreeEnergy_+\left[ \GTfield \right] =&~ \int\limits_{\Omega} \GTFreeEnergy(x,t) I(\GTfield) \mathrm{d}\Omega
\end{align}
A code to simulate Allen-Cahn dynamics and estimate the global observables is provided in
\begin{mdframed}[backgroundcolor=mintedbg, linecolor=mintedbg, innerleftmargin=0, innertopmargin=0,innerbottommargin=0]
\begin{python}
examples.non_local_calculus.Example2_Allen_Cahn.dns
\end{python}
\end{mdframed}

\subsubsection{Reduced order model for Global observables}
\begin{figure}
\centering
\includegraphics[width=0.8\linewidth]{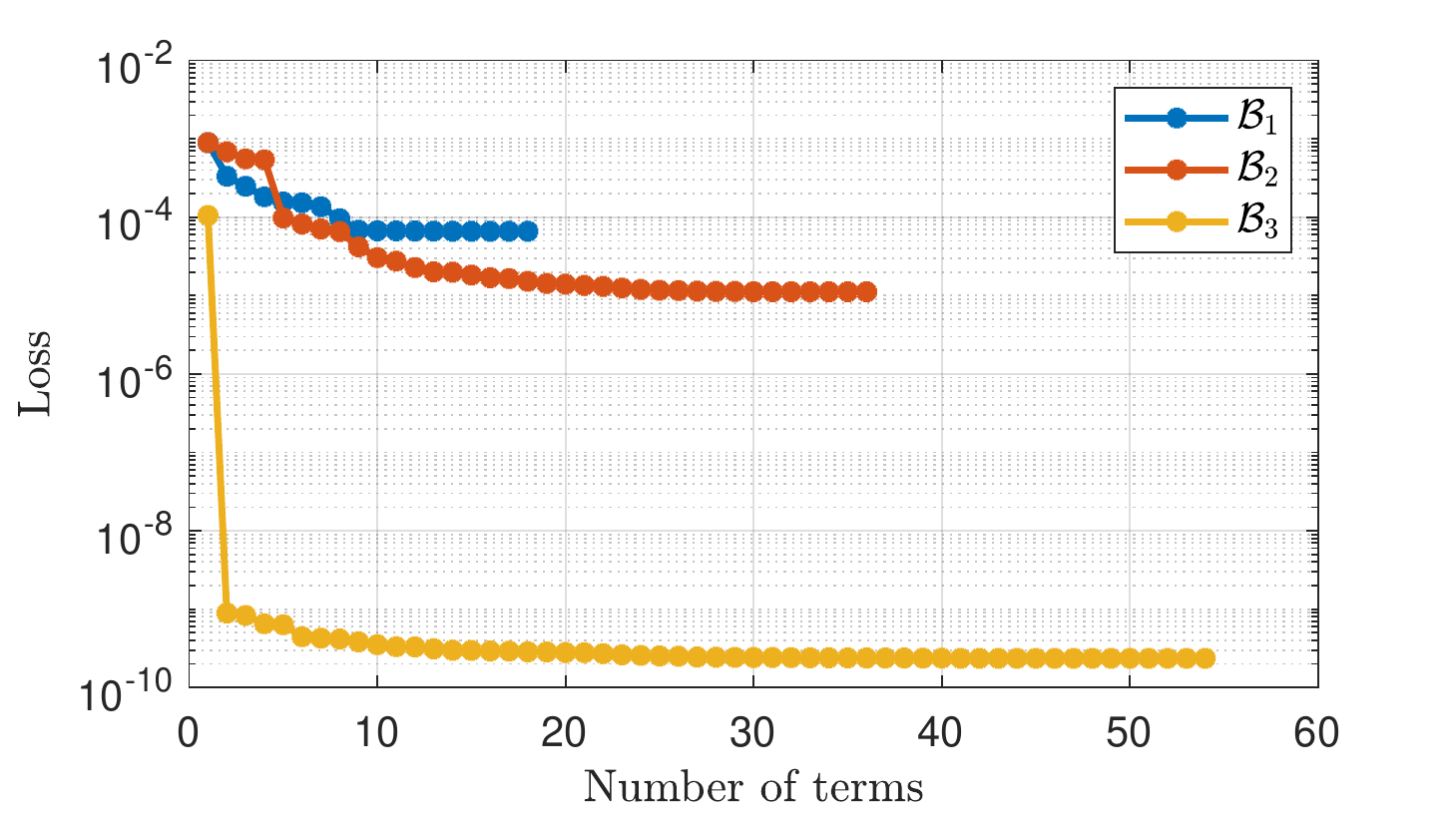}
\caption{Loss during stepwise regression in system identfication of Allen Cahn dynamics.}
\label{fig:allen-cahn-loss}
\end{figure}

We are interested in developing a parsimonious reduced order model for  $\Obsfield_{\phi_+}$. We consider a model for first order kinetics of the observable as follows:
\begin{align}
\frac{d\Obsfield_{\phi_+}}{dt} = \sum_\alpha \gamma^\alpha v_\alpha, \qquad v_\alpha\in \mathcal{B}
\end{align}
Here, $\mathcal{B}$ is the basis set of operators. For demonstration, we choose the following 3 successively expanded basis sets:
\begin{align}
\mathcal{B}_{1} &= \left\lbrace \frac{\delta\Psi_+}{\delta\Obsfield_{g(\GTfield)_+}} \cup \frac{\delta\Psi_+}{\delta\Obsfield_{g(\GTfield)_-}} |  g(\GTfield) \in \mathcal{G}\right\rbrace \\
\mathcal{B}_{2} &= \mathcal{B}_{1} \cup\left\lbrace \frac{\delta\Psi}{\delta\Obsfield_{g(\GTfield)_+}} \cup \frac{\delta\Psi}{\delta\Obsfield_{g(\GTfield)_-}} |  g(\GTfield) \in \mathcal{G}\right\rbrace \\
\mathcal{B}_{3} &= \mathcal{B}_{2} \cup \left\lbrace \Obsfield_{g(\GTfield)_+} \cup \Obsfield_{g(\GTfield)_-} |  g(\GTfield) \in \mathcal{G}\right\rbrace
\end{align}
Our choice of basis sets, $\mathcal{B}_1$ and $\mathcal{B}_2$, is guided by the gradient flow form of the field (Eq.\eqref{eq:ac_gf}). The derived quantities like $\frac{\delta \Psi}{\delta \varphi_{\phi_+}}$ are estimated using the non-local calculus via the code:
\begin{mdframed}[backgroundcolor=mintedbg, linecolor=mintedbg, innerleftmargin=0, innertopmargin=0,innerbottommargin=0]
\begin{python}
examples.non_local_calculus.Example2_Allen_Cahn.estimate_derivative.py
\end{python}
\end{mdframed}
Basis set $\mathcal{B}_3$ is further expanded to include the global observables. For each choice of basis set, a parsimonious model is identified using the stepwise regression strategy via the code:
\begin{mdframed}[backgroundcolor=mintedbg, linecolor=mintedbg, innerleftmargin=0, innertopmargin=0,innerbottommargin=0]
\begin{python}
examples.non_local_calculus.Example2_Allen_Cahn.train_model.py
\end{python}
\end{mdframed}
The training data is provided in terms of 100 trajectories, each estimated with respect to a different initial condition with parameters, $M_{\phi} = 1e-3$, and $ \lambda = 1 $. The resulting coefficients of the trained model are presented in Table~\ref{tab:allen_cahn_model} with the loss curves provided in Fig. \ref{fig:allen-cahn-loss}. We observe that a familiar gradient flow model is recovered as a 1-term model in the case of basis sets, $\mathcal{B}_1$ and $\mathcal{B}_2$. However, in the case of $\mathcal{B}_3$, the loss is considerably lower for models with 2 and more terms. It is also observed that the 2-term model, in case of $\mathcal{B}_3$, approximates an analytical model for this global variable given as  \cite{duschenes2021reduced}:
\begin{align}
\frac{d\Obsfield_{\phi_+}}{dt} = -\GTChemPotential_\GTfield \Obsfield_{f'(\GTfield)_+} \pm \lambda \GTChemPotential_\GTfield \Obsfield_{\Delta\GTfield_\pm} \equiv 4\GTChemPotential_\GTfield \Obsfield_{\GTfield_+} - 4\GTChemPotential_\GTfield \Obsfield_{\GTfield^3_+} \pm \lambda \GTChemPotential_\GTfield \Obsfield_{\Delta\GTfield_\pm}
\end{align}

\begin{table}[]
\centering
\begin{tabular}{|c|c|c|ccccc|}
\hline
\multirow{2}{*}{Basis} & \multicolumn{1}{c|}{\multirow{2}{*}{Iteration}} & \multirow{2}{*}{Loss} & \multicolumn{5}{c|}{Model coefficients} \\ \cline{4-8}
& \multicolumn{1}{c|}{} &  & \multicolumn{1}{c|}{$\gamma^{\frac{\delta\Psi_+}{\delta\varphi_{\phi_+}}}$} & \multicolumn{1}{c|}{$\gamma^{\frac{\delta\Psi_+}{\delta\varphi_{\phi_-}}}$} & \multicolumn{1}{c|}{$\gamma^{\frac{\delta\Psi_+}{\delta\varphi_{f(\phi)_-}}}$} & \multicolumn{1}{c|}{$\gamma^{\frac{\delta\Psi_+}{\delta\varphi_{\phi^3_-}}}$} & $\gamma^{\frac{\delta\Psi_+}{\delta\varphi_{\phi^4_-}}}$ \\ \hline
\multirow{5}{*}{$\mathcal{B}_1$} & 14 & 1.56e-4 & \multicolumn{1}{c|}{-2.81e-1} & \multicolumn{1}{c|}{3.01e-1} & \multicolumn{1}{c|}{-5.84e-2} & \multicolumn{1}{c|}{1.39e-2} & 4.45e-4 \\
& 15 & 1.82e-4 & \multicolumn{1}{c|}{-2.66e-1} & \multicolumn{1}{c|}{2.48e-1} & \multicolumn{1}{c|}{-3.54e-2} & \multicolumn{1}{c|}{4.62e-3} & 0 \\
& 16 & 2.48e-4 & \multicolumn{1}{c|}{-2.68e-1} & \multicolumn{1}{c|}{1.68e-1} & \multicolumn{1}{c|}{-1.18e-2} & \multicolumn{1}{c|}{0} & 0 \\
& 17 & 3.32e-4 & \multicolumn{1}{c|}{-2.90e-1} & \multicolumn{1}{c|}{1.04e-1} & \multicolumn{1}{c|}{0} & \multicolumn{1}{c|}{0} & 0 \\
& 18 & 8.94e-4 & \multicolumn{1}{c|}{-4.36e-1} & \multicolumn{1}{c|}{0} & \multicolumn{1}{c|}{0} & \multicolumn{1}{c|}{0} & 0 \\ \hline
& \multicolumn{1}{c|}{} &  & \multicolumn{1}{c|}{$\gamma^{\frac{\delta\Psi_+}{\delta\varphi_{\phi_+}}}$} & \multicolumn{1}{c|}{$\gamma^{\frac{\delta\Psi}{\delta\varphi_{\phi^2_-}}}$} & \multicolumn{1}{c|}{$\gamma^{\frac{\delta\Psi_+}{\delta\varphi_{f(\phi)_+}}}$} & \multicolumn{1}{c|}{$\gamma^{\frac{\delta\Psi_+}{\delta\varphi_{\phi^2_-}}}$} & $\gamma^{\frac{\delta\Psi_+}{\delta\varphi_{f(\phi)_-}}}$ \\ \hline
\multirow{5}{*}{$\mathcal{B}_2$} & 32 & 9.79e-5 & \multicolumn{1}{c|}{-4.99e-1} & \multicolumn{1}{c|}{-4.48e-1} & \multicolumn{1}{c|}{-2.40e-1} & \multicolumn{1}{c|}{8.49e-1} & 2.02e-1 \\
& 33 & 5.39e-4 & \multicolumn{1}{c|}{-5.61e-1} & \multicolumn{1}{c|}{-6.14e-2} & \multicolumn{1}{c|}{-6.33e-2} & \multicolumn{1}{c|}{4.34e-2} & 0 \\
& 34 & 5.54e-4 & \multicolumn{1}{c|}{-5.51e-1} & \multicolumn{1}{c|}{-2.15e-2} & \multicolumn{1}{c|}{-5.04e-2} & \multicolumn{1}{c|}{0} & 0 \\
& 35 & 6.83e-4 & \multicolumn{1}{c|}{-3.80e-1} & \multicolumn{1}{c|}{-1.79e-2} & \multicolumn{1}{c|}{0} & \multicolumn{1}{c|}{0} & 0 \\
& 36 & 8.94e-4 & \multicolumn{1}{c|}{-4.36e-1} & \multicolumn{1}{c|}{0} & \multicolumn{1}{c|}{0} & \multicolumn{1}{c|}{0} & 0 \\ \hline
& \multicolumn{1}{c|}{} &  & \multicolumn{1}{c|}{$\varphi_{f'(\phi)_+}$} & \multicolumn{1}{c|}{$\varphi_{\Delta\phi_-}$} & \multicolumn{1}{c|}{$\varphi_{\phi_-}$} & \multicolumn{1}{c|}{$\varphi_{\phi^3_+}$} & $\gamma^{\frac{\delta\Psi_+}{\delta\varphi_{\Delta\phi_-}}}$ \\ \hline
\multirow{5}{*}{$\mathcal{B}_3$} & 50 & 6.33e-10 & \multicolumn{1}{c|}{-1.00e+0} & \multicolumn{1}{c|}{-9.90e-4} & \multicolumn{1}{c|}{1.67e-4} & \multicolumn{1}{c|}{-2.10e-4} & -1.04e-3 \\
& 51 & 6.52e-10 & \multicolumn{1}{c|}{-1.00e+0} & \multicolumn{1}{c|}{-9.90e-4} & \multicolumn{1}{c|}{1.28e-4} & \multicolumn{1}{c|}{-1.50e-4} & 0 \\
& 52 & 8.32e-10 & \multicolumn{1}{c|}{-1.00e+0} & \multicolumn{1}{c|}{-9.90e-4} & \multicolumn{1}{c|}{5.31e-5} & \multicolumn{1}{c|}{0} & 0 \\
& 53 & 8.94e-10 & \multicolumn{1}{c|}{-1.00e+0} & \multicolumn{1}{c|}{-9.90e-4} & \multicolumn{1}{c|}{0} & \multicolumn{1}{c|}{0} & 0 \\
& 54 & 1.05e-4 & \multicolumn{1}{c|}{-9.51e-1} & \multicolumn{1}{c|}{0} & \multicolumn{1}{c|}{0} & \multicolumn{1}{c|}{0} & 0 \\ \hline
\end{tabular}
\vspace{2mm}
\caption{The last 5 iterations of stepwise regression in system identfication of Allen Cahn dynamics.}
\label{tab:allen_cahn_model}
\end{table}

\section{Installation, documentation, and contributing}
\label{sec:install}
{
In this section, we briefly discuss the installation steps and documentation, as well as suggest how users may contribute to the library. Users can refer to specific pages of our online documentation for detailed instructions.
}

\subsection{Installation}

\texttt{mechanoChemML} is publicly available on GitHub \cite{mechanoChemML_github} {and is hosted on the Python Package Index (PyPI, https://pypi.org/)}. To install \texttt{mechanoChemML}, we recommend the use of \texttt{Anaconda}. One Can create
a virtual Python package installation environment that is isolated from the Python environment of the operating system with \texttt{Anaconda}.
\begin{mdframed}[backgroundcolor=mintedbg, linecolor=mintedbg, innerleftmargin=0, innertopmargin=0,innerbottommargin=0]
\begin{python}
$(base) conda create --name mechanochemml python==3.7
$(base) conda activate mechanochemml
\end{python}
\end{mdframed}
One can use the following command to install \texttt{mechanoChemML} and its required libraries.
\begin{mdframed}[backgroundcolor=mintedbg, linecolor=mintedbg, innerleftmargin=0, innertopmargin=0,innerbottommargin=0]
\begin{python}
$(mechanochemml) pip install mechanoChemML
\end{python}
\end{mdframed}
To install the proper \texttt{TensorFlow} and \texttt{TensorFlow Probability} version that is compatible with the CUDA version on the user's system, one needs to download the examples provided by the \texttt{mechanoChemML} library via
\begin{mdframed}[backgroundcolor=mintedbg, linecolor=mintedbg, innerleftmargin=0, innertopmargin=0,innerbottommargin=0]
\begin{python}
$(mechanochemml) svn export https://github.com/mechanoChem/mechanoChemML/trunk/examples ./examples
\end{python}
\end{mdframed}
or download the whole library from GitHub via
\begin{mdframed}[backgroundcolor=mintedbg, linecolor=mintedbg, innerleftmargin=0, innertopmargin=0,innerbottommargin=0]
\begin{python}
$(mechanochemml) git clone https://github.com/mechanoChem/mechanoChemML.git mechanoChemML-master
\end{python}
\end{mdframed}
and run the TensorFlow installation script as
\begin{mdframed}[backgroundcolor=mintedbg, linecolor=mintedbg, innerleftmargin=0, innertopmargin=0,innerbottommargin=0]
\begin{python}
$(mechanochemml) python3 examples/install_tensorflow.py
\end{python}
\end{mdframed}
For developers, one can compile the \texttt{mechanoChemML} library and install it locally to reflect the latest GitHub changes that are not available in the released version on the PyPI by the following commands
\begin{mdframed}[backgroundcolor=mintedbg, linecolor=mintedbg, innerleftmargin=0, innertopmargin=0,innerbottommargin=0]
\begin{python}
$(mechanochemml) git clone https://github.com/mechanoChem/mechanoChemML.git mechanoChemML-master
$(mechanochemml) cd mechanoChemML-master/
$(mechanochemml) python3 setup.py bdist_wheel sdist
$(mechanochemml) pip3 install -e .
\end{python}
\end{mdframed}
The newly compiled \texttt{mechanoChemML} library will overwrite the old installed version.

\subsection{Documentation}

The detailed documentation of the \texttt{mechanoChemML} library is provided at \url{https://mechanochemml.readthedocs.io/en/latest/index.html}.
One can use the following commands to further compile a local copy of the documentation files.
\begin{mdframed}[backgroundcolor=mintedbg, linecolor=mintedbg, innerleftmargin=0, innertopmargin=0,innerbottommargin=0]
\begin{python}
$(mechanochemml) cd mechanoChemML-master/docs
$(mechanochemml) make html
\end{python}
\end{mdframed}

\subsection{Contributing}
Instructions for contributing to the \texttt{mechanoChemML} library, such as bug reports, code contribution, documentation contribution, workflow contribution, etc., is discussed at \url{https://mechanochemml.readthedocs.io/en/latest/contribute.html}.
\section{Conclusion}
\label{sec:concl}

Our goal with this communication is to motivate the niche that exists for scientific software between traditional PDE solver libraries and machine learning platforms, and to describe, with an appropriate degree of detail, how the framework of \texttt{mechanoChemML} occupies this position. In addition to code for the machine learning classes, an important idea here is that of machine learning workflows. Laid out with the proper abstraction, these workflows can accommodate a reasonable range of applications in computational materials physics. Moving forward, this library structure will undergo continuous development driven by both: the applications and the evolving understanding of machine learning workflows.

\section{Acknowledgments}

We gratefully acknowledge the support of Toyota Research Institute, Award \#849910: “Computational framework
for data-driven, predictive, multi-scale and multi-physics modeling of battery materials”. This work has also been supported in part by National Science Foundation DMREF grant \#1729166, ``Integrated Framework for Design of Alloy-Oxide Structures''.
Additional support was provided by Defense Advanced Research Projects Agency (DARPA) under Agreement No. HR0011199002, ``Artificial Intelligence guided multi-scale multi-physics framework for discovering complex emergent materials phenomena''. Computing resources were
provided in part by the National Science Foundation, United States via grant 1531752 MRI: Acquisition of Conflux,
A Novel Platform for Data-Driven Computational Physics (Tech. Monitor: Ed Walker). This work also used the
Extreme Science and Engineering Discovery Environment (XSEDE) Comet at the San Diego Supercomputer Center
and Stampede2 at The University of Texas at Austin’s Texas Advanced Computing Center through allocation TGMSS160003
and TG-DMR180072.

\section*{Data availability}
The raw data required to reproduce these findings are available to download from \url{https://github.com/mechanoChem/mechanoChemML}. The processed data required to reproduce these findings are available to download from \url{https://github.com/mechanoChem/mechanoChemML}.

\bibliographystyle{unsrtnat}

\end{document}